\documentclass[a4paper,12pt]{article}

\usepackage{amssymb,amstext,amsmath}
\usepackage{tikz}

\newcommand{\bra}[1]{{\langle{#1}\vert}}
\newcommand{\ket}[1]{{\vert{#1}\rangle}}
\newcommand{\bracket}[2]{\langle #1 \vert #2 \rangle}
\newcommand{\Tr}{\mathop{\rm Tr}\nolimits}
\newcommand{\realni}{\ensuremath{\mathbb{R}}}
\newcommand{\del}{\partial}

\newcommand{\orto}{\bot}
\newcommand{\nablalr}{\stackrel{\leftrightarrow}{\nabla}\!\!{}}
\newcommand{\nablar}{\stackrel{\rightarrow}{\nabla}\!\!{}}
\newcommand{\nablal}{\stackrel{\leftarrow}{\nabla}\!\!{}}
\newcommand{\lc}{\varepsilon}

\newcommand{\ds}{\displaystyle}
\newcommand{\tst}{\textstyle}

\newcommand{\rmd}{{\rm d}}

\newcommand{\cC}{{\cal C}}
\newcommand{\cD}{{\cal D}}
\newcommand{\cH}{{\cal H}}
\newcommand{\cK}{{\cal K}}
\newcommand{\cL}{{\cal L}}
\newcommand{\cM}{{\cal M}}
\newcommand{\cN}{{\cal N}}

\begin{document}

\title{\bf Gauge protected entanglement between gravity and matter}

\author{Nikola Paunkovi\'c$^{1,2,3}$ and Marko Vojinovi\'c$^{4}$}

\date{}

\maketitle

\begin{center}
\small {\it
$^1$Instituto de Telecomunica\~coes, Avenida Rovisco Pais 1, 1049-001 Lisbon, Portugal \\
$^2$Departamento de Matem\'atica, Instituto Superior T\'ecnico, Universidade de Lisboa, \\ Avenida Rovisco Pais 1, 1049-001 Lisbon, Portugal \\
$^3$CeFEMA, Instituto Superior T\'ecnico, Universidade de Lisboa, \\ Avenida Rovisco Pais 1, 1049-001 Lisbon, Portugal \\
$^4$Institute of Physics, University of Belgrade, \\ Pregrevica 118, 11080 Belgrade, Serbia}

\bigskip

E-mail: \texttt{npaunkov@math.tecnico.ulisboa.pt}, \texttt{vmarko@ipb.ac.rs}
\end{center}

\begin{abstract}
We show that, as a consequence of the local Poincar\'e symmetry, gravity and matter fields have to be entangled, unless the overall action is carefully fine-tuned. First, we present a general  argument, applicable to any particular theory of quantum gravity with matter, by performing the analysis in the abstract nonperturbative canonical framework, demonstrating the nonseparability of the scalar constraint, thus promoting the entangled states as the physical ones. Also, within the covariant framework, using a particular toy model, we show explicitly that the Hartle-Hawking state in the Regge model of quantum gravity is entangled. Our result is potentially relevant for the quantum-to-classical transition, taken within the framework of the decoherence programme: due to the gauge symmetry requirements, the matter does not decohere, it is by default decohered by gravity. Generically, entanglement is a consequence of interaction. This new entanglement could potentially, in form of an ``effective interaction'', bring about corrections to the weak equivalence principle, further confirming that spacetime as a smooth four-dimensional manifold is an emergent phenomenon. Finally, the existence of the gauge-protected entanglement between gravity and matter could be seen as a criterion for a plausible theory of quantum gravity, and in the case of perturbative quantisation approaches, a confirmation of the persistence of the manifestly broken gauge symmetry.
\end{abstract}


\section{\label{SecIntroduction}Introduction}

The unsolved problems of formulating quantum theory of gravity (QG) and interpreting quantum mechanics (QM) are arguably the two most prominent ones of the modern theoretical physics. So far, most of the approaches to solve the two were studied independently. Indeed, the majority of the interpretations of QM do not involve explicit dynamical effects (with notable exceptions of the spontaneous collapse and the de Broglie-Bohm theories), while the researchers from the QG community often just adopt some particular interpretation of QM, assuming that it contains no unresolved issues. Nevertheless, the two problems share a number of similar unsolved questions and counter-intuitive features. A prominent example is nonlocality: entanglement-based nonlocality in QM, as well as the anticipated explicit dynamical nonlocality in QG (a consequence of quantum superpositions of different gravitational fields, i.e., different spacetimes and their respective causal orders). Another prominent issue relevant for both standard QM and QG is the quantum-to-classical transition and the related measurement problem.

In relation to the latter, decoherence is in QM the standard approach to the emergence of classicality: due to huge complexity of macroscopic (``classical'') systems and the surrounding environment (bath), the ({\em for all practical purposes}) inevitable  interaction between the two leads to the entanglement and the loss of coherence. While technically this is completely within the standard QM, when coupled with additional assumptions, such as the many-world interpretation (likely to be the predominant within the community working on decoherence and quantum-to-classical transition), the decoherence offers a possible solution to the measurement problem. In an alternative approach, problems with quantising gravity led to the half century old idea of gravitationally induced objective collapse of the wave function~\cite{kar:66} (for an overview, see for example~\cite{bas:loc:sat:sin:bulb:13}, chapter III.B): roughly speaking, due to the position uncertainty of massive bodies, which are the sources of gravitational field, the latter exhibits quantum fluctuations that decohere the matter, forcing it (or, rather both the matter and gravity) to collapse in a well defined (classical) state. Without invoking objective collapse, decoherence of quantum matter by purely classical gravity was studied in \cite{BruknerNatureCom,BruknerNaturePhys}. In the context of perturbative quantum gravity, the topic of gravitationally induced decoherence of matter, taken purely within the scope of standard QM (i.e., in the same fashion in which macroscopic bodies decohere due to inevitable interaction with surrounding photons, neutrinos, microwave background radiation, etc.), became recently an intensive field of research~\cite{oni:wan:16}, see also~\cite{liu:sou:wang:16} and the references therein for decoherence in the context of cosmological inflation. In addition, a lot of research focuses on entanglement induced by the presence of horizons in curved spacetime, in approaches based on the holography conjecture and in the studies of the black hole information problem~\cite{don:16} (for a review, see recent lecture notes~\cite{raa:16}). In particular, these approaches study the entanglement between the degrees of freedom ({\em both} gravitational and matter) on the two sides of the horizon.

In this paper we study the entanglement between gravitational and matter fields, in the context of an abstract nonperturbative theory of quantum gravity, as well as on the example of the Hartle-Hawking state in the Regge quantum gravity model, and show that the two fields should {\em always} be entangled. Our approach is different from the standard one, studied in the perturbative framework: instead of ``for all practical purposes'' inevitable fast interaction-induced decoherence from initially product states between two sub-systems \cite{oni:wan:16,kay:98,kok:yur:03,ham:mar:llo:car:sev:mar:10,ham:mar:11,bas:gro:usb:17}, we show that the gauge symmetry requirements (coming in particular from the local Poincar\'e symmetry) secure the entangled states between matter and gravity as physical states. We call the latter the {\em gauge-protected} decoherence, in contrast to the {\em dynamical} decoherence of the former. In addition, unlike the horizon-based studies, we discuss the entanglement between the gravitational and the matter degrees of freedom, rather than between the two specially chosen regions of spacetime.

Our analysis rests on two main assumptions. First, we assume the validity of the local Poincar\'e symmetry at the quantum level. In the classical field theory, the local Poincar\'e symmetry is a formalisation of the principle of general relativity, which is one of the foundational principles of Einstein's theory of gravity. It is therefore natural to assume that this gauge symmetry exists at the quantum level as well. Second, at the classical level we assume the validity of the equivalence principle, which is also the main ingredient of Einstein's general relativity. In particular, we assume its ``strong'' version, namely that the equivalence principle applies to all matter fields (i.e., all non-gravitational fields) present in nature.

Given these two assumptions, we focus on the general nonperturbative abstract canonical quantisation of the gravitational and matter fields, thus giving a generic model-independent argument for a theory of quantum gravity with matter. We analyse the consequences of the local Poincar\'e symmetry-enforced scalar, 3-diffeomorphism and local Lorentz constraints on the structure of the total Hilbert space of the theory. Namely, since the physical states must be invariant with respect to the gauge symmetry, the constraints induce the Gupta-Bleuler-like conditions on the state vectors. Based on the equivalence principle, we then show that the particular non-separable form of the scalar constraint renders typical product states non-invariant. Thus, it eliminates the product states from the physical Hilbert space of the quantum theory, unless the interaction between gravity and matter is specifically designed to circumvent the non-invariance of product states. In this way, the local Poincar\'e symmetry protects the existence of entanglement between the gravitational and matter fields.

In order to verify our results obtained within the abstract canonical framework, we also study the covariant (i.e., path integral) quantisation. In particular, knowing that the Hartle-Hawking state~\cite{har:haw:83} satisfies the scalar constraint, and is therefore an element of the physical Hilbert space, we explicitly test whether the matter and gravitational fields are entangled for this state vector. We perform the calculation in the Regge quantum gravity model, since it is one of the simplest models which provide an explicit definition of the gravitational path integral with matter, and show that the gravitational and matter fields are indeed entangled for the Hartle-Hawking state constructed on a simple toy example triangulation.

Therefore, our analysis shows that either gravity and matter fields are indeed entangled, or there exists an additional, unknown property of the action, implementing the fine tuning needed to allow for the invariance of separable states.

The paper is organised as follows. Section~\ref{SecGravConstraint} is divided into three subsections. The first is devoted to the recapitulation of the Hamiltonian structure of Poincar\'e gauge theories. The second outlines the procedure of nonperturbative canonical quantisation of constrained systems and its application to the case of gravity with matter fields. In the third subsection we use those results to show that the scalar constraint suppresses the existence of separable states of a matter-gravity system. In section~\ref{SecDensityMatrix}, we present a standard entanglement criterion for pure bipartite quantum states and discuss it, within the framework of the path integral quantisation, for the case of the Hartle-Hawking state of quantum fields of gravity and matter. In section~\ref{SecRQGexample}, we first introduce the Regge model of quantum gravity, and then apply it to evaluate the entanglement criterion for the Hartle-Hawking state, demonstrating that gravity and matter are indeed entangled in this state. Finally, in section~\ref{SecConclusions} we present the summary of the results, their discussion, and possible future lines of research.

It is important to stress that the gauge-protected entanglement is not an automatic consequence of the universal coupling between gravity and matter, or the fact that matter fields are always defined over some background spacetime geometry. For example, in perturbative gravity approach, it is quite possible to write the separable state between gravity and matter as
$$
\ket{\Psi} = \ket{g} \otimes \ket{\phi}\,,
$$
where $\ket{g}$ is the graviton state vector, while $\ket{\phi}$ is the state vector of a scalar particle (both vectors obtained by acting with graviton and scalar creation operators on the Minkowski vacuum state $\ket{0} \equiv \ket{0}_G \otimes \ket{0}_M$). The reason why such a state can be considered legitimate is that local Poincar\'e symmetry is explicitly broken in the perturbative gravity approach, with both matter and gravity being treated as spin-zero and spin-two fields, respectively, living on a Minkowski spacetime manifold. A similar situation arises in perturbative string theory, where local Poincar\'e symmetry is also manifestly broken. However, in quantum gravity models where the local Poincar\'e symmetry is not violated, our analysis shows that a generic product state between gravity and matter would fail to be gauge invariant. Thus, the gauge-protected entanglement between gravity and matter is a nontrivial statement and a consequence of local Poincar\'e symmetry, rather than an automatic property of matter fields living on a spacetime manifold.

Our notation and conventions are as follows. We will work in the natural system of units in which $c=\hbar=1$  and $G = l_p^2$, where $l_p$ is the Planck length. By convention, the metric of spacetime will have the spacelike Lorentz signature $(-,+,+,+)$. The spacetime indices are denoted with lowercase Greek letters $\mu,\nu,\dots$ and take the values $0,1,2,3$. The spatial part of these, taking values $1,2,3$, will be denoted with lowercase Latin letters $i,j,\dots$ from the middle of the alphabet. The $SO(3,1)$ group indices will be denoted with the lowercase Latin letters $a,b,\dots$ from the beginning of the alphabet, and take the values $0,1,2,3$. The Lorentz-invariant metric tensor is denoted as $\eta_{ab}$. The capital Latin indices $A,B,\dots$ count the field components in a particular representation of the $SO(3,1)$ group, and take the values from $1$ up to the dimension of that representation. Quantum operators will always carry a hat, $\hat{\phi}(x)$, $\hat{g}(x)$, etc. Finally, we will systematically denote the values of functions with parentheses, $f(x)$, while functionals will be denoted with brackets, $F[\phi]$.

\section{\label{SecGravConstraint}Entanglement from the scalar constraint}


This section is dedicated to the analysis of the constraints imposed by the relativity and equivalence principles. In subsection~\ref{SubSecClassicalHamiltonianStructure} we briefly recapitulate the classical Hamiltonian structure of gravitational interaction, followed by a short review of canonical quantisation, presented in subsection~\ref{SubSecCanonicalQuantization}. After that, in subsection~\ref{SubSecMainArgument} we present the main result of our paper: we show that the scalar constraint, and possibly the 3-diffeomorphism constraint, bring about the generic entanglement between gravity and matter.

\subsection{\label{SubSecClassicalHamiltonianStructure}Hamiltonian structure of Poincar\'e gauge theories}

We begin with a short review of the Hamiltonian structure of gravitational interaction, based on the local Poincar\'e symmetry. This subsection is aimed to be only a review of the main results, so we will skip all proofs and derivations. The details of the Hamiltonian structure for Poincar\'e gauge theories (PGT) can be found in many textbooks, see for example \cite{bla:02}, chapter V, and the references therein.

We will assume a foliation of spacetime into space and time, with the spacetime topology $\cM_4 = \Sigma_3 \times \realni$, where $\Sigma_3$ is the $3D$ hypersurface. For the purpose of generality, we will describe the gravitational field as $g(x)$ and matter fields as $\phi(x)$, without specifying their exact field content, except in examples. A typical example would be the Einstein-Cartan gravity coupled to a Dirac matter field, so that the choice of fundamental gravitational fields $g$ would be the tetrads  $e^a{}{}_{\mu}(x)$ and the spin connection $\omega^{ab}{}_{\mu}(x)$, while the choice for the fundamental matter field $\phi$ would be a Dirac fermion field $\psi(x)$. However, other choices for $g$ and $\phi$ are also possible, for example the metric tensor $g_{\mu\nu}$ for gravity and the electromagnetic potential $A^{\mu}$ for matter, etc. Since our analysis is largely independent of such choices, we will stick to the abstract notation $g$ and $\phi$, assuming that one can apply our analysis to each particular concrete choice of fundamental fields.

Given the above notation, we will assume that the action of the theory can be written as
\begin{equation} \label{UkupnoOpsteDejstvo}
S[g,\phi] = S_G[g] + S_M[g,\phi]\,,
\end{equation}
where $S_G[g]$ is the action of the pure gravitational field, while $S_M[g,\phi]$ is the action of the matter fields coupled to gravity. Since the spacetime metric must both be a function of the gravitational field $g$ and is always present in the definition of the dynamics of matter fields, the action for the matter fields cannot contain terms independent of $g$. This elementary fact is the crux of our main argument below, and is justified by the equivalence principle, which dictates how matter couples to gravity.

To a large extent, we also do not need to specify the details of the actions $S_G[g]$ and $S_M[g,\phi]$. We will only assume that the action (\ref{UkupnoOpsteDejstvo}) belongs to the PGT class of theories, i.e., that it is invariant with respect to local Poincar\'e group $P(4) = \realni^4 \ltimes SO(3,1)$. Every theory belonging to the PGT class has the Hamiltonian with the following general structure \cite{bla:02}:
\begin{equation} \label{Hamiltonian}
H = \int_{\Sigma_3} \rmd^3 \vec{x} \left[ N \cC + N^i \cC_i + N^{ab} \cC_{ab} \right]\,,
\end{equation}
up to a $3$-divergence. Here $N$, $N^i$ and $N^{ab}$ are Lagrange multipliers, the first two of which are commonly known as lapse and shift functions. The quantities $\cC$, $\cC_i$ and $\cC_{ab}$ are usually known as the scalar constraint, $3$-diffeomorphism constraint, and the local Lorentz constraint (sometimes also called the Gauss constraint), respectively. They are a $(g,\phi)$-field representation of the $10$ generators of the Poincar\'e group $P(4)$, in particular the time translation generator, the three space translation generators, and six local Lorentz generators (rotations and boosts). Note that the Hamiltonian (\ref{Hamiltonian}) is always a linear combination of these constraints.

The constraints in (\ref{Hamiltonian}) have the structure similar to the structure of the gravity-matter action (\ref{UkupnoOpsteDejstvo}), namely
\begin{equation} \label{StrukturaConstraintova}
\begin{array}{lcl}
\cC & = & \ds \cC^G(g,\pi_g) + \cC^M(g,\pi_g,\phi,\pi_{\phi})\,, \\
\cC_i & = & \ds \cC^G_i(g,\pi_g) + \cC^M_i(g,\pi_g,\phi,\pi_{\phi})\,, \\
\cC_{ab} & = & \ds\cC^G_{ab}(g,\pi_g) + \cC^M_{ab}(g,\pi_g,\phi,\pi_{\phi})\,, \\
\end{array}
\end{equation}
where $\pi_g$ and $\pi_{\phi}$ are the momenta canonically conjugated to the fields $g$ and $\phi$, respectively, defined as functional derivatives of the action with respect to the time-derivatives of the fields,
$$
\pi_g(x) = \frac{\delta S}{\delta \del_0 g(x)}\,, \qquad \pi_{\phi}(x) =  \frac{\delta S}{\delta \del_0 \phi(x)}\,.
$$
The general dependence (\ref{StrukturaConstraintova}) on the fields and momenta reflects the corresponding dependence in (\ref{UkupnoOpsteDejstvo}).

The exact forms of the gravitational terms of the constraints, namely $\cC^G$, $\cC^G_i$ and $\cC^G_{ab}$, will be immaterial for our main argument presented in the subsection \ref{SubSecMainArgument} below. In contrast, the structure of the matter terms $\cC^M$, $\cC^M_i$ and $\cC^M_{ab}$ will be crucial, so we discuss it here in more detail. Choose a matter field such that it transforms according to some specific irreducible transformation of the Poincar\'e group, and denote it as $\phi^A(x)$, where the capital index $A$ counts the field components in that representation. Then the $3$-diffeo constraint $\cC^M_i$ and the Gauss constraint $\cC^M_{ab}$ are given as
\begin{equation} \label{FormOfKinematicConstraints}
\cC^M_i (g,\pi_g,\phi,\pi_{\phi}) = \pi_{\phi A} \nabla_i{}^A{}_B \phi^B\,, \qquad
\cC^M_{ab} (g,\pi_g,\phi,\pi_{\phi}) = \pi_{\phi A} (M_{ab})^A{}_B \phi^B\,,
\end{equation}
where $\nabla_i{}^A{}_B$ is a covariant derivative for the irreducible representation according to which the field $\phi$ transforms, while $(M_{ab})^A{}_B$ is the representation of the generator $M_{ab}$ of the Lorentz group $SO(3,1)$ in the same representation. In general, the covariant derivative depends on the spacetime metric or connection, which is a function of the fundamental gravitational fields $g$, and possibly their momenta $\pi_g$. The Lorentz group generators, on the other hand, do not depend on the spacetime geometry, so the Gauss constraint is actually independent of $g$ and $\pi_g$, and we can write $\cC^M_{ab} (g,\pi_g,\phi,\pi_{\phi}) = \cC^M_{ab} (\phi,\pi_{\phi})$.

In order to illustrate the two constraints, we will write (\ref{FormOfKinematicConstraints}) for the scalar and Dirac fields, as the most elementary examples. In the case of the scalar field, we write $\phi^A(x) = \varphi(x)$, where the index $A$ takes only a single value. The covariant derivative acts on the scalar field as an ordinary derivative, while the representation of the Lorentz generators is trivial, so we can write
\begin{equation} \label{FormOfKinematicConstraintsForScalarField}
\cC^M_i (\varphi,\pi_{\varphi}) = \pi_{\varphi} \del_i \varphi\,, \qquad
\cC^M_{ab} (\varphi,\pi_{\varphi}) = \pi_{\varphi} \varphi\,.
\end{equation}
We see that in the case of the scalar field, both constraints are independent of the gravitational fields and their momenta. In the case of the Dirac fields, we write $\phi^A(x) = (\psi^A(x),\bar{\psi}^A(x))$, where the index $A$ now represents the spinorial index, and we will omit writing it. The covariant derivative acts on the Dirac field in the standard way,
\begin{equation} \label{ActionOfNablaOnSpinors}
\nablar_{\mu} \psi \equiv \nabla_{\mu} \psi \equiv \del_{\mu} \psi + \frac{1}{2} \omega^{ab}{}_{\mu} \sigma_{ab} \psi \,, \qquad
\bar{\psi} \nablal_{\mu}\equiv \del_{\mu}\bar{\psi} - \frac{1}{2} \omega^{ab}{}_{\mu} \bar{\psi} \sigma_{ab} \,,
\end{equation}
where $\omega^{ab}{}_{\mu}$ is the spin connection, $\sigma_{ab} = \frac{1}{4} [\gamma_a,\gamma_b]$, and $\gamma_a$ are the standard Dirac gamma-matrices satisfying the anticommutation relation $\{\gamma_a,\gamma_b\} = -2\eta_{ab}$. The representation of the Lorentz generators for the case of the Dirac field is $M_{ab} = \sigma_{ab}$. Denoting the conjugate momentum for $\psi$ as $\bar{\pi}$ and conjugate momentum for $\bar{\psi}$ as $\pi$, we can write the constraints (\ref{FormOfKinematicConstraints}) as:
\begin{equation} \label{FormOfKinematicConstraintsForDiracField}
\cC^M_i (\omega,\psi,\bar{\pi},\bar{\psi},\pi) = \bar{\pi} \nabla_i \psi + (\bar{\psi} \nablal_i) \pi\,, \qquad
\cC^M_{ab} (\psi,\bar{\pi},\bar{\psi},\pi) = \bar{\pi} \sigma_{ab} \psi - \bar{\psi} \sigma_{ab} \pi\,.
\end{equation}
Note that here, unlike in the scalar field example, the $3$-diffeo constraint contains the spin connection $\omega^{ab}{}_{\mu}$, which is a part of the gravitational field $g = (e^a{}_{\mu},\omega^{ab}{}_{\mu})$ for the Einstein-Cartan gravity.

In contrast to the $3$-diffeo and Gauss constraints (\ref{FormOfKinematicConstraints}), the scalar constraint $\cC^M$ has a more complicated form,
\begin{equation} \label{FormOfScalarConstraint}
\cC^M(g,\pi_g,\phi,\pi_{\phi}) = \pi_{\phi A} \nabla_{\orto}{}^A{}_B \phi^B - \frac{1}{N} \cL_M(g,\pi_g,\phi,\pi_{\phi})\,,
\end{equation}
where the matter Lagrangian density is defined via
$$
S_M[g,\phi] = \int \rmd^4x\, \cL_M(g,\del g,\phi,\del\phi)\,,
$$
and $\nabla_{\orto} \equiv n^{\mu} \nabla_{\mu}$ is the covariant derivative in the direction of the timelike vector $n^{\mu}$ orthogonal to the spacelike hypersurface $\Sigma_3$. The vector $n^{\mu}$ obviously depends on the spacetime metric $g_{\mu\nu}$, and is thus a function of the fundamental gravitational fields $g$.

There are several things to note regarding the scalar constraint (\ref{FormOfScalarConstraint}). First, it is clear that $N\cC^M$ is the Legendre transformation of the Lagrangian density $\cL_M$ with respect to the ``velocity'' $N \nabla_{\orto} \phi$. Second, in contrast to the constraints (\ref{FormOfKinematicConstraints}), which depend only on the symmetry transformation properties of the fields, the form of the scalar constraint (\ref{FormOfScalarConstraint}) depends also on the choice of the matter Lagrangian density $\cL_M$, and is therefore described by the dynamics of the matter fields coupled to gravity. And third, the scalar constraint $\cC^M$ always necessarily depends on the gravitational fields $g$, in contrast to the $3$-diffeo constraint which may or may not depend on $g$, and the Gauss constraint which never depends on $g$. As we already suggested above, this is because the Lagrangian of the matter fields coupled to gravity always contains the gravitational degrees of freedom, courtesy of the equivalence principle.

Let us illustrate this dependence of $\cC^M$ on the gravitational fields $g$ in the case of the Dirac field. The action for the Dirac field $\phi = (\psi,\bar{\psi})$ coupled to the gravitational fields $g = (e^a{}{}_{\mu},\omega^{ab}{}_{\mu})$ is given as
\begin{equation} \label{DiracFieldAction}
S_M[e,\omega,\psi,\bar{\psi}] = \int \rmd^4x\, e \left( \frac{i}{2} \bar{\psi} \gamma^a e^{\mu}{}_a \nablalr_{\mu} \psi - m \bar{\psi} \psi \right)\,,
\end{equation}
where $e$ is the determinant of the tetrad $e^a{}_{\mu}$, while $e^{\mu}{}_a$ is the inverse tetrad. In addition, $\nablalr_{\mu} \equiv \nablar_{\mu} - \nablal_{\mu}$, and the covariant derivatives $\nablar_{\mu}$ and $\nablal_{\mu}$ act to the right and to the left as defined in (\ref{ActionOfNablaOnSpinors}), from which one can see that the action also explicitly depends on the connection $\omega^{ab}{}_{\mu}$. From the action one can read off the Lagrangian density, and calculate the scalar constraint (\ref{FormOfScalarConstraint}) as
$$
\cC^M(e,\omega,\psi,\bar{\psi}) = -\frac{e}{N} \left( \frac{i}{2} \bar{\psi} \gamma^a e^{\mu}{}_a \left( \delta^{\mu}_{\nu} + n^{\mu} n_{\nu} \right) \nablalr_{\nu} \psi - m \bar{\psi} \psi \right) \,.
$$
Note that the quantity $\delta^{\mu}_{\nu} + n^{\mu} n_{\nu}$ is a projector to the hypersurface $\Sigma_3$.

\subsection{\label{SubSecCanonicalQuantization}Canonical quantisation}

Having discussed the Hamiltonian structure of the action (\ref{UkupnoOpsteDejstvo}), we now pass on to a short description of the canonical quantisation of the theory. The quantisation of an arbitrary physical system with constraints is performed in the standard way, using the Dirac's procedure \cite{DirProc:58,DirProc:64} (see \cite{bla:02} for a review). One begins by classifying all constraints of the theory into the first and the second class. The second class constraints are then eliminated by passing from the Poisson brackets to the Dirac brackets. The first class constraints remain and represent the generators of the gauge symmetry. In general, the Hamiltonian of the theory can be written as
\begin{equation} \label{GenericHamiltonianWithConstraints}
H = H_0 + \lambda^A \cC_A\,,
\end{equation}
where $\lambda^A$ are Lagrange multipliers, $\cC_A$ are first class constraints, and $H_0$ is the part of the Hamiltonian which describes the evolution of the physical degrees of freedom. Given all this, the quantisation is performed in the Heisenberg picture, promoting fundamental fields $\phi(x)$ to quantum mechanical operators $\hat{\phi}(x)$, and introducing the state vectors $\ket{\Psi} \in \cH_{\rm kin}$, where $\cH_{\rm kin}$ is the kinematical Hilbert space of the theory. The Dirac brackets between the fields and their momenta are then promoted to the commutators of the corresponding operators. The Hamiltonian, being a functional of the fields and momenta, also becomes an operator, providing the usual Heisenberg equations of motion for the field operators,
$$
i \frac{\del \hat{\phi}(x)}{\del t} = [\hat{\phi}(x),\hat{H}]\,.
$$
Finally, the kinematical Hilbert space $\cH_{\rm kin}$ is projected onto its gauge invariant subspace $\cH_{\rm phys}$, by requiring that every state vector $\ket{\Psi} \in \cH_{\rm phys}$ is annihilated by the generators of the gauge symmetry group,
$$
\hat{\cC}_A \ket{\Psi} = 0\,.
$$
In quantum electrodynamics these conditions are known as Gupta-Bleuler quantisation conditions \cite{gup:50,ble:50}. This requirement ensures that the gauge symmetry of the classical theory remains to be a symmetry of the quantum theory as well.

Of course, one cannot hope to implement the above quantisation programme in full detail for the general action (\ref{UkupnoOpsteDejstvo}), especially without the detailed specification of the fundamental degrees of freedom that define the theory. Instead, we assume that the quantisation programme has been carried out in detail, and that all quantities we will write are well defined. This approach has one important feature and one important drawback. The feature is generality --- our main argument for the inevitable entanglement between gravity and matter, to be presented in subsection \ref{SubSecMainArgument}, should hold for every particular quantum theory constructed in the above way, as it does not actually depend on the details of the quantisation. The drawback is abstractness --- in using such a general formalism and making a flat assumption that all details are well defined, we lose the capability to provide any concrete examples. That said, in section \ref{SecRQGexample} we discuss one rigorously defined example of a theory of quantum gravity with matter (Regge quantum gravity), and demonstrate the entanglement between gravity and matter fields. Unlike the canonical quantisation discussed in this section, that example will be done in the framework of the path integral quantisation.

Keeping this disclaimer in mind, we proceed along the lines outlined above and perform the canonical quantisation. The most prominent property of our model is the structure of the Hilbert space of the theory. The initial kinematical Hilbert space $\cH_{\rm kin}=\cH_G \otimes \cH_M$ has a natural product structure between the gravitational and matter Hilbert spaces, since we have two sets of fields, $\hat{g}$ and $\hat{\phi}$, corresponding to gravity and matter, respectively. Thus, we have a naturally preferred bipartite physical system, because gravitational and matter degrees of freedom can be fully distinguished from each other. Second, in order to preserve the Poincar\'e gauge symmetry of the theory at the quantum level, we have to pass from the kinematical Hilbert space to the gauge invariant, physical Hilbert space $\cH_{\rm phys}$. By definition, a state $\ket{\Psi} \in \cH_{\rm kin}$ is an element of $\cH_{\rm phys}$ iff it satisfies
\begin{equation} \label{QuantumKinematicConstraintEquations}
\begin{array}{ccl}
\hat{\cC}_{ab} \ket{\Psi} & \equiv & \ds \left[ \cC^G_{ab}(\hat{g},\hat{\pi}_g) + \cC^M_{ab}(\hat{\phi},\hat{\pi}_{\phi}) \right] \ket{\Psi} = 0\,, \\
\hat{\cC}_i \ket{\Psi} & \equiv & \ds \left[ \cC^G_i(\hat{g},\hat{\pi}_g) + \cC^M_i(\hat{g},\hat{\pi}_g,\hat{\phi},\hat{\pi}_{\phi}) \right] \ket{\Psi} = 0\,, \\
\end{array}
\end{equation}
and
\begin{equation} \label{QuantumScalarConstraintEquation}
\begin{array}{ccl}
\hat{\cC} \ket{\Psi} & \equiv & \ds \left[ \cC^G(\hat{g},\hat{\pi}_g) + \cC^M(\hat{g},\hat{\pi}_g,\hat{\phi},\hat{\pi}_{\phi}) \right] \ket{\Psi} = 0\,. \\
\end{array}
\end{equation}
As stated above, we assume that the operators $\hat{\cC}_{ab}$, $\hat{\cC}_i$ and $\hat{\cC}$ are well defined, that operator ordering choice has been fixed, as well as all other necessary technical choices, in order for the expressions above to make sense mathematically. 

We argue that, due to these constraint equations, there are no states in $\cH_{\rm phys}$ which can be written as product states of the form $\ket{\Psi_G} \otimes \ket{\Psi_M}$, where $\ket{\Psi_G}\in \cH_G$ and $\ket{\Psi_M} \in \cH_M$, i.e., the states in $\cH_{\rm phys}$ are entangled. We focus on the scalar constraint (\ref{QuantumScalarConstraintEquation}), while the constraints (\ref{QuantumKinematicConstraintEquations}) are either irrelevant or redundant for our analysis. This main argument of our paper is presented in the next subsection.

\subsection{\label{SubSecMainArgument}Entanglement}

Given a state vector $\ket{\Psi} \in \cH_{\rm kin} = \cH_G \otimes \cH_M$, it is an element of the physical Hilbert space $\cH_{\rm phys}$ if it satisfies the Gauss and $3$-diffeo constraints (\ref{QuantumKinematicConstraintEquations}) and the scalar constraint (\ref{QuantumScalarConstraintEquation}). Choosing the eigenbases $\{ \ket{g} \}$ and $\{ \ket{\phi} \}$ of the quantum field operators $\hat{g}$ and $\hat{\phi}$, respectively, we can work in the so-called field representation, defined as
\begin{equation} \label{FieldRepresentationDefinition}
\bra{g} \hat{g} = g \bra{g}\,, \qquad \bra{g} \hat{\pi}_g = -i \frac{\delta}{\delta g} \bra{g}\,, \qquad
\bra{\phi} \hat{\phi} = \phi \bra{\phi}\,, \qquad \bra{\phi} \hat{\pi}_{\phi} = -i \frac{\delta}{\delta \phi} \bra{\phi}\,.
\end{equation}
Acting on (\ref{QuantumScalarConstraintEquation}) with $\bra{g,\phi} \equiv \bra{g} \otimes \bra{\phi}$ from the left, the scalar constraint becomes a functional partial differential equation of Wheeler-DeWitt type:
\begin{equation} \label{FunctionalQuantumScalarConstraintEquation}
\tst \left[ \cC_G\left(g,-i\frac{\delta}{\delta g} \right) + \cC_M\left(g,-i\frac{\delta}{\delta g} ,\phi,-i\frac{\delta}{\delta \phi} \right) \right] \Psi[g,\phi] = 0\,,
\end{equation}
where $\Psi[g,\phi] \equiv \bracket{g,\phi}{\Psi}$ is the wavefunctional of the combined gravity-matter system. We now try to look for a separable state, in the form $\ket{\Psi} = \ket{\Psi_G} \otimes \ket{\Psi_M}$, where $\ket{\Psi_G} \in \cH_G$ and $\ket{\Psi_M}\in \cH_M$, as a solution of this equation. Using the field representation (\ref{FieldRepresentationDefinition}), we write the wavefunctional $\Psi[g,\phi]$ as
\begin{equation} \label{ProductState}
\Psi[g,\phi] \equiv \bracket{g,\phi}{\Psi} = \left( \bra{g}\otimes \bra{\phi} \right) \left( \ket{\Psi_G} \otimes \ket{\Psi_M} \right) = \bracket{g}{\Psi_G} \bracket{\phi}{\Psi_M} \equiv \Psi_G[g] \Psi_M[\phi]\,.
\end{equation}
Equation (\ref{FunctionalQuantumScalarConstraintEquation}) can have separable solutions $\Psi[g,\phi] = \Psi_G[g] \Psi_M[\phi]$ if the functional differential operator $\cC_M$ can be written as a product of two operators, denoted $\cK_G$ and $\cK_M$, depending only on $(g,\frac{\delta}{\delta g})$ and on $(\phi,\frac{\delta}{\delta\phi})$, respectively,
\begin{equation} \label{SeparabilityCriterion}
\tst \cC_M\left(g,-i\frac{\delta}{\delta g},\phi,-i\frac{\delta}{\delta \phi} \right) = \cK_G\left( g,\frac{\delta}{\delta g} \right) \cK_M \left( \phi,\frac{\delta}{\delta \phi} \right)\,.
\end{equation}
If such operators $\cK_G$ and $\cK_M$ exist so that (\ref{SeparabilityCriterion}) holds, the scalar constraint equation (\ref{FunctionalQuantumScalarConstraintEquation}) can be rewritten as
$$
\tst \Psi_M[\phi]\; \cC_G\left( g,-i\frac{\delta}{\delta g} \right) \Psi_G[g] = - \left[ \cK_G\left( g,\frac{\delta}{\delta g} \right) \Psi_G[g]\right] \left[ \cK_M \left( \phi,\frac{\delta}{\delta \phi} \right) \Psi_M[\phi] \right] \,.
$$
Dividing this with $\Psi_M[\phi]\; \cK_G\left( g,\frac{\delta}{\delta g} \right) \Psi_G[g]$, assuming it is well-defined, we obtain
$$
\frac{1}{\cK_G\left( g,\frac{\delta}{\delta g} \right) \Psi_G[g]}\, \cC_G\left( g,-i{\tst\frac{\delta}{\delta g}} \right) \Psi_G[g] = - \frac{1}{\Psi_M[\phi]} \, \cK_M \left( \phi,{\tst\frac{\delta}{\delta \phi}} \right) \Psi_M[\phi] = A\,,
$$
where $A$ is a constant, since the terms on the left and the right of the first equality depend on different sets of variables. Therefore, the above equation splits into two independent equations,
\begin{equation} \label{DekuplovaneSvojstveneJednacineZaOperatoreK}
\tst \left[ \cC_G\left( g,-i\frac{\delta}{\delta g} \right) -A \; \cK_G\left( g,\frac{\delta}{\delta g} \right) \right] \Psi_G[g] = 0 \,, \qquad
\left[ \cK_M \left( \phi,\frac{\delta}{\delta \phi} \right) + A \right] \Psi_M[\phi] = 0 \,,
\end{equation}
which are to be solved independently for $\Psi_G[g]$ and $ \Psi_M[\phi]$, thus providing a separable solution of (\ref{FunctionalQuantumScalarConstraintEquation}).

The whole procedure above rests on the assumption (\ref{SeparabilityCriterion}) that the matter part $\cC_M$ of the scalar constraint operator can be written as a product of two operators $\cK_G$ and $\cK_M$. Our main argument is to demonstrate that the assumption (\ref{SeparabilityCriterion}) is never satisfied for the usual matter fields, due to the universal nature of the coupling of gravity to matter, ultimately dictated by the equivalence principle. Namely, given the structure of the classical scalar constraint for matter (\ref{FormOfScalarConstraint}), the corresponding operator can be written as
\begin{equation} \label{FormOfQuantumScalarConstraint}
\cC^M(\hat{g},\hat{\pi}_g,\hat{\phi},\hat{\pi}_{\phi}) = \hat{\pi}_{\phi A} \hat{\nabla}_{\orto}{}^A{}_B \hat{\phi}^B - \frac{1}{N} \cL_M(\hat{g},\hat{\pi}_g,\hat{\phi},\hat{\pi}_{\phi})\,,
\end{equation}
where a certain ordering of the operators is assumed. The constraint (\ref{FormOfQuantumScalarConstraint}) features the operator-valued matter Lagrangian $\cL_M$. Therefore, in order to demonstrate that $\cC_M$ does not satisfy the separability criterion (\ref{SeparabilityCriterion}) it is enough to demonstrate that the matter Lagrangian does not satisfy it. This can be done on a case-by-case basis, for each particular matter field. Invoking the equivalence principle, we can write the operator-valued Lagrangian for the scalar field coupled to gravity as
$$
\cL_M (\hat{g},\hat{\varphi},\del\hat{\varphi}) = \frac{1}{2} \hat{e} \left[ \hat{g}^{\mu\nu} (\del_{\mu} \hat{\varphi}) (\del_{\nu} \hat{\varphi}) - m^2 \hat{\varphi}^2 + U(\hat{\varphi}) \right]\,,
$$
where $\hat{e}$ is the square-root of the minus determinant operator of the metric tensor,
$$
\hat{e} \equiv \left[ \frac{1}{4!}\lc^{\alpha\beta\gamma\delta} \lc^{\mu\nu\rho\sigma} \hat{g}_{\alpha\mu} \hat{g}_{\beta\nu} \hat{g}_{\gamma\rho} \hat{g}_{\delta\sigma} \right]^{\frac{1}{2}}\,,
$$
and $U$ is some interaction potential of the scalar field. Ignoring the multiplicative factor $\hat{e}$ that acts only on $\cH_G$, the Lagrangian is a sum of two types of terms: the kinetic term, containing the inverse metric $\hat{g}^{\mu\nu}$, and the mass and potential terms not featuring the gravitational field in any form. The sum cannot therefore be factored into the form $\cK_G(\hat{g})\cK_M(\hat{\phi},\del\hat{\phi})$, since the Lagrangian is not a homogeneous function of the gravitational degrees of freedom. Even in the case of the massless free scalar field, i.e., when $m=0$ and $U=0$, the kinetic term is a sum of several different components of the metric and the derivatives of the scalar field,
$$
\hat{g}^{00} (\del_0 \hat{\varphi}) (\del_0 \hat{\varphi}) + \hat{g}^{01} (\del_0 \hat{\varphi}) (\del_1 \hat{\varphi}) + \hat{g}^{12} (\del_1 \hat{\varphi}) (\del_2 \hat{\varphi}) + \dots
$$
and this still cannot be factored into a product of two operators $\cK_G$ and $\cK_M$.

In the case of the Dirac field, again invoking the equivalence principle, the operator-valued Lagrangian is given by (\ref{DiracFieldAction}),
$$
\cL_M(\hat{e},\hat{\omega},\hat{\psi},\hat{\bar{\psi}}) = \hat{e} \left( \frac{i}{2} \hat{\bar{\psi}} \gamma^a \hat{e}^{\mu}{}_a \hat{\nablalr}_{\mu} \hat{\psi} - m \hat{\bar{\psi}} \hat{\psi} \right)\,.
$$
Like in the case of the scalar field, the kinetic and mass terms in the Lagrangian depend differently on the gravitational fields $\hat{e}^a{}_{\mu}$ and $\hat{\omega}^{ab}{}_{\mu}$, and $\cL_M$ cannot be factored. Moreover, the kinetic term itself cannot be factored, since it is a sum of two terms (see equations (\ref{ActionOfNablaOnSpinors})), only one of which contains the spin connection $\hat{\omega}^{ab}{}_{\mu}$.

Next, the operator-valued Lagrangian for the electromagnetic field coupled to gravity has the form
$$
\cL_M(\hat{g},\hat{A},\del\hat{A}) = - \frac{1}{4} \hat{e} \, \hat{g}^{\mu\rho} \hat{g}^{\nu\sigma} \hat{F}_{\mu\nu} \hat{F}_{\rho\sigma}\,,
$$
where $\hat{F}_{\mu\nu} \equiv \del_{\mu}\hat{A}_{\nu} - \del_{\nu} \hat{A}_{\mu}$. Applying the same argument as in the case of the free massless scalar field, this Lagrangian also cannot be factored into the form $\cK_G \cK_M$. The same argument also applies to the case of the non-Abelian Yang-Mills Lagrangians.

Summing up, given the ways the matter fields are coupled to gravity, based on the equivalence principle, we conclude that the separability criterion (\ref{SeparabilityCriterion}) is never satisfied for the physically relevant cases of scalar, spinor and vector fields. Therefore, according to the discussion above, the scalar constraint (\ref{QuantumScalarConstraintEquation}) should not admit separable state vectors into $\cH_{\rm phys}$.

Regarding the above analysis, it is important to emphasize the following. Namely, one should note that it is in principle possible for equation (\ref{FunctionalQuantumScalarConstraintEquation}) to have product state solutions (\ref{ProductState}) despite the fact that it does not satisfy the separability criterion (\ref{SeparabilityCriterion}). In other words, the criterion (\ref{SeparabilityCriterion}) is a sufficient condition for the existence of product state solutions of (\ref{FunctionalQuantumScalarConstraintEquation}), but it is not necessary, so its violation does not strictly imply the absence of product state solutions. Nevertheless, given the arguably highly complex structure of equation (\ref{FunctionalQuantumScalarConstraintEquation}) --- meaning that it represents a nonlinear functional partial differential equation of at least second order in $g$ and $\phi$ --- it is natural to regard any potential product state solutions as completely accidental. Moreover, it is questionable if the boundary conditions required for such solutions correspond to any realistic physical situation in nature, i.e., they could be irrelevant for realistic physics. Due to all these arguments, the existence of product state solutions, in spite of the violation of the separability criterion (\ref{SeparabilityCriterion}), is in our opinion an extraordinary claim, and as such requires extraordinary evidence. In other words, the burden of proof is in fact with the statement that any product state solution exists, rather than the opposite. Consequently, product states (\ref{ProductState}) are generically not elements of $\cH_{\rm phys}$, and even if one can prove that there exist some product states which do happen to belong to $\cH_{\rm phys}$, such states would arguably be completely accidental, with questionable relevance for physics. Otherwise, if there exists a whole class of separable states which solve (\ref{FunctionalQuantumScalarConstraintEquation}) despite the violation of the criterion (\ref{SeparabilityCriterion}), there must be some deep eluding property of the scalar constraint equation, which is both completely unknown and very interesting to study.

Finally, while it turns out that the analysis of the scalar constraint equation (\ref{QuantumScalarConstraintEquation}) is sufficient for our conclusions, let us briefly mention the status of the remaining two constraint equations (\ref{QuantumKinematicConstraintEquations}). First, the Gauss constraint $\hat{\cC}_{ab}$ obviously admits separable state vectors. On the other hand, the situation with the $3$-diffeo constraint $\hat{\cC}_i$ is more complicated, and the conclusion depends on the type of the field. For example, in the case of the scalar field, from (\ref{FormOfKinematicConstraintsForScalarField}) we read that $\hat{\cC}^M_i$ depends only on the scalar field and its momentum, which means that the constraint equation does admit separable state vectors. However, in the case of the Dirac field, from (\ref{FormOfKinematicConstraintsForDiracField}) we read that $\hat{\cC}^M_i$ depends on the spin connection in addition to the Dirac field, and this dependence is not homogeneous in the spin connection, see (\ref{ActionOfNablaOnSpinors}). Thus, the $3$-diffeo constraint equation does not admit separable state vectors. However, the behaviour of the Gauss and $3$-diffeo constraint equations is redundant for our argument, since the scalar constraint equation (\ref{QuantumScalarConstraintEquation}) already suppresses separable state vectors for all fields, due to the dynamical form of the coupling of matter to gravity. Therefore, our initial assumption of local Poincar\'e symmetry can be weakened to the localisation of its translational subgroup, while the generators of the local Lorentz subgroup are irrelevant for our argument.


\section{\label{SecDensityMatrix} Entanglement in the path integral framework}

In the previous section we have discussed the gauge-protected entanglement within the framework of the canonical quantisation of the gravitational field with matter. In this section, we focus instead on the path integral framework of quantisation. We analyse the entanglement on the example of the Hartle-Hawking state, which is known to satisfy all constraints of the theory. In the next section, we are going to apply the results of this section to the concrete case of Regge quantum gravity.

First, we discuss an entanglement criterion for the case of pure overall state of the gravity and matter fields. We begin with a brief recapitulation of basic results from the standard QM and quantum information theory. A pure bipartite state $\ket{\Psi}_{12} \in \mathcal{H}_1 \otimes \mathcal{H}_2$ of systems $1$ and $2$ can be written in the Schmidt bi-orthogonal form (see, for example~\cite{nie:chu:04}):
\begin{equation}
\label{schmidt}
\ket{\Psi}_{12} = \sum_i \sqrt{r_i} \ket{\alpha_i}_1 \otimes \ket{\beta_i}_2,
\end{equation}
where $\{ \ket{\alpha_i}_1 \}$ and $\{ \ket{\beta_i}_2 \}$ are two sets of mutually orthogonal states from $\mathcal{H}_1$ and $\mathcal{H}_2$, respectively. The partial sub-system states are then given as
\begin{equation}
\label{rho_1}
\hat\rho_{1} = \sum_i r_i \ket{\alpha_i}_1 \otimes \bra{\alpha_i}_1,
\end{equation}
for the system $1$, and analogously for the system $2$. Squaring $\hat\rho_{1}$, we have
\begin{equation}
\hat\rho_{1}^2 = \sum_i r_i^2 \ket{\alpha_i}_1 \otimes \bra{\alpha_i}_1.
\end{equation}
If the overall state $\ket{\Psi}_{12}$ is separable (i.e., a simple product state), the above sum in~(\ref{rho_1}) will be trivial, consisting of a single projector onto the ray $\ket{\alpha_1}_1 \otimes \bra{\alpha_1}_1$, with $r_1 = 1$. Thus, we have that $\hat{\rho}_1^2 = \hat{\rho}_1$, or simply, $\Tr \hat{\rho}_1^2 = \Tr \hat{\rho}_1 = 1$. In case the state $\ket{\Psi}_{12}$ is entangled, the sum~(\ref{rho_1}) will consist of more than just one term, resulting in $(\forall \  i) \ r_i < 1$. Therefore, $(\forall \  i) \ r_i^2 < r_i$, and we finally have
\begin{equation}
\label{ent_crit}
\Tr \hat\rho_{1}^2 = \sum_i r_i^2 < \sum_i r_i = \Tr \hat\rho_{1} = 1.
\end{equation}
Due to the symmetry of the Schmidt form~(\ref{schmidt}), the same is valid for the system $2$ (for the formal proof of the above entanglement criterion~(\ref{ent_crit}), see for example~\cite{nie:chu:04}).

After this recapitulation of the standard results from QM, we proceed with the analysis of the bipartite system of the gravity (G) and matter (M) fields, applying the above entanglement criterion~(\ref{ent_crit}) to the case of quantum fields. For simplicity, we omit the subscripts G and M for pure states of gravity and matter, respectively.

Let $\cH_{\rm kin}=\cH_G \otimes \cH_M$ be the combined kinematical gravity-matter Hilbert space. Denote the bases in $\cH_G$ and $\cH_M$ as $\{ \ket{g} \}$ and $\{ \ket{\phi} \}$, respectively. These are the eigenbases of the corresponding quantum field operators $\hat{g}$ and $\hat{\phi}$, evaluated on the $3D$ boundary $\Sigma_3 = \del \cM_4$ of the $4D$ spacetime manifold $\cM_4$. The general state vector $\ket{\Psi} \in \cH_{\rm kin}$ of the gravity-matter system can then be written as
\begin{equation} \label{GeneralWavefunctional}
\ket{\Psi} = \int \cD g \int \cD \phi \, \Psi[g,\phi] \, \ket{g} \otimes \ket{\phi}\,,
\end{equation}
where $\Psi[g,\phi]=\bracket{g,\phi}{\Psi}$ is called the wavefunctional (in analogy to wavefunction from quantum mechanics), and the functional integrals over gravitational degrees of freedom $g$ and matter degrees of freedom $\phi$ are assumed to be well defined in some way (in section \ref{SecRQGexample} we present an explicit example of this). The bases $\{ \ket{g} \}$ and $\{ \ket{\phi} \}$ are assumed to be orthonormal, satisfying
\begin{equation} \label{OrthogonalityRelations}
\bracket{g}{g'} = \delta[g-g']\,, \qquad \bracket{\phi}{\phi'} = \delta[\phi-\phi']\,,
\end{equation}
where the Dirac delta functional is assumed to satisfy the formal functional integral identities
\begin{equation} \label{DiracDeltaIntegrals}
\int \cD g \, F[g] \delta[g-g'] = F[g']\,, \qquad \int \cD \phi \, F[\phi] \delta[\phi-\phi'] = F[\phi']\,,
\end{equation}
for any functionals $F[g]$ and $F[\phi]$ belonging to some suitable relevant class.

From the state (\ref{GeneralWavefunctional}) one can construct a reduced density matrix $\hat{\rho}_M$ for matter fields, by taking the partial trace over gravitational degrees of freedom of the full density matrix $\hat{\rho} \equiv \ket{\Psi} \otimes \bra{\Psi}$, as
$$
\hat{\rho}_M = \Tr_G \hat{\rho} = \int \cD g \, \bra{g} \Big( \ket{\Psi} \otimes \bra{\Psi} \Big) \ket{g}\,.
$$
Substituting (\ref{GeneralWavefunctional}) we get
$$
\hat{\rho}_M = \int \cD g \int \cD g' \int \cD \phi' \int \cD g'' \int \cD \phi'' \, \Psi^*[g',\phi'] \Psi[g'',\phi'']\, \bra{g} \Big( \ket{g''} \otimes \ket{\phi''} \otimes \bra{g'} \otimes \bra{\phi'} \Big) \ket{g}\,.
$$
Using (\ref{OrthogonalityRelations}) and (\ref{DiracDeltaIntegrals}), the expression for the reduced density matrix can be evaluated to
\begin{equation} \label{IzrazZaRhoM}
\hat{\rho}_M = \int \cD g \int \cD \phi' \int \cD \phi'' \, \Psi^*[g,\phi'] \Psi[g,\phi'']\, \ket{\phi''} \otimes \bra{\phi'} \,.
\end{equation}
Taking the square and using (\ref{OrthogonalityRelations}) and (\ref{DiracDeltaIntegrals}) again, one obtains
$$
\hat{\rho}_M^2 = \int \cD g \int \cD g' \int \cD \phi' \int \cD \phi'' \int \cD \phi''' \, \Psi^*[g,\phi'] \Psi[g,\phi''] \Psi^*[g',\phi'''] \Psi[g',\phi'] \, \ket{\phi''} \otimes \bra{\phi'''} \,.
$$
Finally, taking the trace over matter fields,
$$
\Tr_M \hat{\rho}_M^2 = \int \cD \phi \, \bra{\phi} \hat{\rho}_M^2 \ket{\phi}\,,
$$
we get
\begin{equation} \label{OpstiIzrazZaTrRhoSq}
\Tr_M \hat{\rho}_M^2 = \int \cD g \int \cD g' \int \cD \phi \int \cD \phi' \, \Psi^*[g,\phi'] \Psi[g,\phi] \Psi^*[g',\phi] \Psi[g',\phi'] \,.
\end{equation}

Now we want to evaluate (\ref{OpstiIzrazZaTrRhoSq}) for one specific state, namely the Hartle-Hawking state, denoted $\ket{\Psi_{\rm HH}}$. This state is known to satisfy the scalar constraint equation (\ref{QuantumScalarConstraintEquation}), see  \cite{har:haw:83}, and thus belongs to the physical Hilbert space $\cH_{\rm phys}$. Our aim is to demonstrate that the Hartle-Hawking state is nonseparable, and the strategy is to argue that $\Tr_M \hat{\rho}_M^2 < 1$ for $\hat{\rho} = \ket{\Psi_{\rm HH}} \otimes \bra{\Psi_{\rm HH}}$. The Hartle-Hawking state is defined by specifying the wavefunctional $\Psi[g,\phi]$ in (\ref{GeneralWavefunctional}) as
\begin{equation} \label{DefinicijaHartleHawkingStanja}
\Psi_{\rm HH}[g,\phi] = \cN \int \cD G \int \cD \Phi \, e^{i S_{\rm tot}[g,\phi,G,\Phi]} \,.
\end{equation}
Here $\cN$ is a normalisation constant, the variables $G$ and $\Phi$ (denoted with the capital letters) live in the bulk spacetime $\cM_4$, while $g$ and $\phi$ (denoted with lowercase letters) live on the boundary $\Sigma_3 = \del \cM_4$, as before. The path integrals are taken over the bulk while keeping the boundary fields constant. Finally, the total action functional $S_{\rm tot}$ has the following structure
\begin{equation} \label{DefTotDejstva}
S_{\rm tot}[g,\phi,G,\Phi] = S_G[g,G] + S_M[g,\phi,G,\Phi]\,,
\end{equation}
where $S_G$ is the action for the gravitational field (for example the Einstein-Hilbert action with a cosmological constant), while $S_M$ is the action for the matter fields coupled to gravity --- hence its dependence on both the gravitational and matter fields. See \cite{har:haw:83} for details on the construction of the expression (\ref{DefinicijaHartleHawkingStanja}).

In order to analyse the expression (\ref{OpstiIzrazZaTrRhoSq}) more efficiently, it is convenient to introduce the following quantity,
\begin{equation} \label{DefinicijaZ}
Z[\phi,\phi'] \equiv \int \cD g \, \Psi_{\rm HH}[g,\phi] \Psi_{\rm HH}^*[g,\phi']\,,
\end{equation}
which represents the matrix element of the reduced density matrix $\hat{\rho}_M$. Namely, by evaluating (\ref{IzrazZaRhoM}) for the Hartle-Hawking state, one obtains
\begin{equation} \label{ZkaoMatricniElementOdRho}
\hat{\rho}_M = \int \cD \phi \int \cD \phi' \, Z[\phi,\phi']\, \ket{\phi} \otimes \bra{\phi'} \,.
\end{equation}
In addition, $Z[\phi,\phi']$ has an important geometric structure. Namely, one can consider two copies of the spacetime manifold $\cM_4$, where the boundary $\Sigma_3$ of the first copy features the fields $g,\phi$, while the boundary of the second copy features the fields $g,\phi'$, i.e., such that the gravitational field $g$ is the same, while matter fields $\phi$ and $\phi'$ are different on the boundaries. Then one takes the second copy of $\cM_4$, inverts it with respect to the boundary $\Sigma_3$ (the result is denoted as $\bar{\cM}_4$), and glues it to the first copy along the common boundary, to obtain a manifold $\cM_4 \cup \bar{\cM}_4$, which has no boundary. This can be illustrated by the following diagrams:
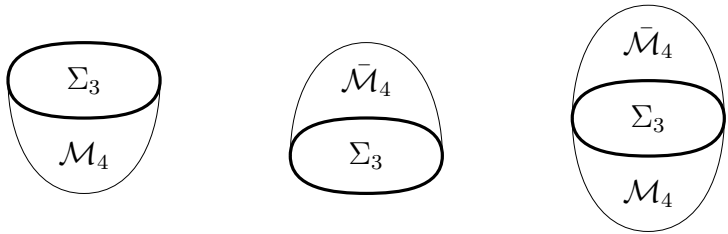
\begin{center}
\begin{tabular}{ccccc}
\begin{tikzpicture}
\draw[very thin] (0,1.5) to [out=270, in=180 ] (1,0) to [out=0, in=270] (2,1.5);
\draw[very thick]               (0,1.5) to [out=270, in=180 ] (1,1) to [out=0, in=270] (2,1.5) to [out=90, in=0] (1,2) to [out=180, in=90] (0,1.5);
\node at (1,0.5) {$\cM_4$};
\node at (1,1.5) {$\Sigma_3$};
\filldraw[white] (1,-0.5) circle (0pt);
\end{tikzpicture}
& $\qquad$ &
\begin{tikzpicture}
\draw[very thin] (0,1.5) to [out=90, in=180 ] (1,3) to [out=0, in=90] (2,1.5);
\draw[very thick]               (0,1.5) to [out=270, in=180 ] (1,1) to [out=0, in=270] (2,1.5) to [out=90, in=0] (1,2) to [out=180, in=90] (0,1.5);
\node at (1,2.5) {$\bar{\cM}_4$};
\node at (1,1.5) {$\Sigma_3$};
\filldraw[white] (1,0.5) circle (0pt);
\end{tikzpicture}
& $\qquad$ &
\begin{tikzpicture}
\draw[very thin] (0,1.5) to [out=270, in=180 ] (1,0) to [out=0, in=270] (2,1.5);
\draw[very thin] (0,1.5) to [out=90, in=180 ] (1,3) to [out=0, in=90] (2,1.5);
\draw[very thick]               (0,1.5) to [out=270, in=180 ] (1,1) to [out=0, in=270] (2,1.5) to [out=90, in=0] (1,2) to [out=180, in=90] (0,1.5);
\node at (1,0.5) {$\cM_4$};
\node at (1,2.5) {$\bar{\cM}_4$};
\node at (1,1.5) {$\Sigma_3$};
\end{tikzpicture}
\\
\end{tabular}
\end{center}
The quantity $Z[\phi,\phi']$ is then obtained by integrating over all gravitational degrees of freedom, and all bulk matter degrees of freedom, weighted by the kernel $e^{iS_{\rm tot}}$ of the Hartle-Hawking wavefunction (\ref{DefinicijaHartleHawkingStanja}). This construction is important because the trace of $Z[\phi,\phi']$ is the state sum of the gravitational and matter fields over the manifold $\cM_4 \cup \bar{\cM}_4$:
\begin{equation} \label{VezaStateSumeIZ}
\int \cD \phi \, Z[\phi,\phi] = Z \equiv \int \cD G \int \cD \Phi \, e^{iS[G,\Phi]}\,.
\end{equation}
Here, $S[G,\Phi]$ is the total gravity-matter action similar to (\ref{DefTotDejstva}), defined over the manifold $\cM_4 \cup \bar{\cM}_4$, and thus features no boundary fields. From (\ref{DefinicijaZ}), (\ref{ZkaoMatricniElementOdRho}) and (\ref{VezaStateSumeIZ}) it is then easy to see that the normalization of the state sum, $Z=1$, and simultaneously the normalization of the reduced density matrix, $\Tr \hat{\rho}_M = 1$, i.e.,
\begin{equation} \label{NormalizacijaZ}
\int \cD \phi \, Z[\phi,\phi] = 1\,,
\end{equation}
are equivalent to the normalisation of the Hartle-Hawking state, $\bracket{\Psi_{\rm HH}}{\Psi_{\rm HH}} = 1$. Finally, from the definition (\ref{DefinicijaZ}) it is easy to see that $Z[\phi,\phi']$ is self-adjoint,
$$
Z[\phi,\phi'] = Z^*[\phi',\phi]\,,
$$
as the matrix elements of the density matrix $\hat{\rho}_M$ are supposed to be.

Returning to the evaluation of (\ref{OpstiIzrazZaTrRhoSq}) for the Hartle-Hawking state, one can use (\ref{DefinicijaZ}) to rewrite it into the compact form
\begin{equation} \label{TrRhoSqPrekoZ}
\Tr_M \hat{\rho}_M^2 = \int \cD \phi \int \cD \phi' \, \left| Z[\phi,\phi'] \strut \right|^2 \,.
\end{equation}
At this point the general analysis cannot proceed any further, since the right-hand side cannot be evaluated explicitly without specifying the details of the theory. The calculation will therefore proceed further in the next section, where we consider one detailed model of quantum gravity with matter.

Despite the inability to evaluate the integral (\ref{TrRhoSqPrekoZ}) in the general case, one can give a qualitative argument that the result is not equal to one, leading to the nonseparability of the Hartle-Hawking state. Namely, given the definition (\ref{DefinicijaHartleHawkingStanja}) of the Hartle-Hawking state, it is easy to see that it essentially depends on two quantities --- the normalisation constant $\cN$, and the choice of the action $S_{\rm tot}$. The normalisation constant is fixed by the requirement that (\ref{NormalizacijaZ}) holds. This leaves the value of the integral (\ref{TrRhoSqPrekoZ}) depending solely on the choice of the classical action of the theory. It is qualitatively straightforward to see that different choices of the action will lead to different values of $\Tr_M \hat{\rho}_M^2$, so any generic choice of $S_{\rm tot}$ is likely to give $\Tr_M \hat{\rho}_M^2 < 1$. A tentative choice for (\ref{DefTotDejstva}) would be the Einstein-Hilbert action for $S_G$ and the Standard Model of elementary particle physics for $S_M$, based on the gauge group $SU(3) \times SU(2) \times U(1)$. However, we know that the Standard Model action is incomplete, for example due to the fact that dark matter is not included in the description. Therefore, the choice of the classical action is a sort of a moving target, and it is unlikely that any candidate action we choose will give $\Tr_M \hat{\rho}_M^2 = 1$. In this sense, one can only conclude that in a generic case the Hartle-Hawking state is nonseparable, supporting the abstract argument from section \ref{SecGravConstraint}.

Finally, let us note that our assumption of local Poincar\'e gauge symmetry implies that we are discussing the Lorentzian path integral formulation of the theory. In contrast, within the Euclidean approach, the Hartle-Hawking state has some problematic characteristics, see for example \cite{FLT} and references therein.

\section{\label{SecRQGexample}Regge quantum gravity example}

In this section we will present a short review of the Regge quantum gravity model coupled to scalar matter, and then use this model to evaluate (\ref{TrRhoSqPrekoZ}) for the Hartle-Hawking state. The Regge quantum gravity model is intimately connected to the covariant loop quantum gravity research framework \cite{rov:04,rov:vid:14}, its generalisations \cite{mik:voj:12,mik:13,voj:16}, and various related research areas \cite{Hamber,Loll} (see also \cite{FL} for an interesting connection to the noncommutative geometry approach in the $3D$ case). Nevertheless, it can be introduced and studied as a simple standalone model of quantum gravity independent of any other context, as was done in \cite{pau:voj:16}, where some preliminary results regarding the entanglement in the Hartle-Hawking state have been announced.

\subsection{\label{SubSecRQGFormalism}Formalism of Regge quantum gravity}

The Regge quantum gravity model is arguably the simplest toy-model of quantum gravity constructed by providing a rigorous definition for the gravitational path integral, generically denoted as
\begin{equation} \label{OpstiGravitacioniPathIntegral}
Z_G = \cN \int \cD g \, e^{iS_{EH}[g]}\,,
\end{equation}
where $S_{EH}[g]$ is the Einstein-Hilbert action for general relativity. The construction of the path integral follows Feynman's original idea of path integral definition-by-discretisation. We begin by passing from a smooth $4D$ spacetime manifold $\cM_4$ to a piecewise-linear $4D$ manifold, most commonly a triangulation $T(\cM_4)$. This structure naturally features $4$-simplices $\sigma$ as basic building blocks, which themselves consist of tetrahedra $\tau$, triangles $\Delta$, edges $\epsilon$ and vertices $v$. The invariant quantities associated to these objects are the $4$-volume of the $4$-simplex ${}^{(4)}V_{\sigma}$, the $3$-volume of the tetrahedron ${}^{(3)}V_{\tau}$, the area of the triangle $A_{\Delta}$ and the length of the edge $l_{\epsilon}$, respectively, while the vertices do not have nontrivial quantities assigned to them.

It is important to emphasise that the edge lengths are most fundamental of all these quantities, since one can always uniquely express ${}^{(4)}V_{\sigma}$, ${}^{(3)}V_{\tau}$ and $A_{\Delta}$ as functions of $l_{\epsilon}$. For example, the most well-known is the Heron formula for the area of a triangle in terms of its three edge lengths,
$$
A_{\Delta}(l) = \sqrt{s(s-l_1)(s-l_2)(s-l_3)}\,, \qquad s\equiv \frac{l_1+l_2+l_3}{2}\,,
$$
where the three edges $\epsilon=1,2,3$ belong to the triangle $\Delta$.

Given a spacetime triangulation, the Einstein-Hilbert action of general relativity,
$$
S_{EH}[g] = -\frac{1}{16\pi l_p^2} \int_{\cM_4} \rmd^4x \sqrt{-g}\, R(g)\,,
$$
can be reformulated in terms of edge lengths of the triangulation as the Regge action
$$
S_R[l] = -\frac{1}{8\pi l_p^2} \sum_{\Delta\in T(\cM_4)} A_{\Delta}(l) \delta_{\Delta}(l)\,, 
$$
where $\delta_{\Delta}$ is the so-called deficit angle at triangle $\Delta$, measuring the amount of spacetime curvature around $\Delta$. See \cite{Regge} and \cite{Hamber} for details and a review.

Once the classical action for general relativity has been adapted to a piecewise-linear manifold structure, we can take the edge lengths of the edges in the triangulation as the fundamental degrees of freedom of the theory, and define the gravitational path integral (\ref{OpstiGravitacioniPathIntegral}) as:
\begin{equation} \label{RQGgravStateSumaDef}
Z_G = \cN \int_D \prod_{\epsilon\in T(\cM_4)} \rmd l_{\epsilon} \, \mu(l) e^{iS_R[l]}\,.
\end{equation}
Here $\cN$ is a normalisation constant, while $\mu(l)$ is the measure term which ensures the convergence of the state sum $Z_G$. For the purpose of this paper, we choose the exponential measure
\begin{equation} \label{IzabranaRQGmera}
\mu(l) = \exp \left( -\frac{1}{L_{\mu}^4} \sum_{\sigma\in T(\cM_4)} {}^{(4)}V_{\sigma}(l) \right)\,,
\end{equation}
where $L_{\mu}>0$ is a constant and a free parameter of the model (see \cite{Mikovic,MikovicVojinovicCCletter,MikovicVojinovicCCveliki} for motivation and analysis). Note that the sum of the $4$-volumes of all $4$-simplices gives the total $4$-volume of the triangulation $T(\cM_4)$, and will sometimes be denoted simply as $V_4$.

The choice of edge lengths as the fundamental gravitational degrees of freedom in (\ref{RQGgravStateSumaDef}) determines the integration domain $D$ as a subset of the Cartesian product $(\realni_0^+)^E$, where $E$ is the total number of edges in $T(\cM_4)$, while $\realni_0^+$ is the maximum integration domain of each individual edge length. We should note that $D$ is a strict subset of $(\realni_0^+)^E$ due to the presence of triangle inequalities which must be satisfied for all triangles, tetrahedra and $4$-simplices in a given triangulation.

Once we have defined the gravitational path integral (\ref{OpstiGravitacioniPathIntegral}) via the state sum (\ref{RQGgravStateSumaDef}), it is straightforward to generalise this definition to the situation which includes matter fields. For simplicity, we will discuss only a single real scalar field $\varphi$, although it is not a problem to include other fields as well. The path integral we are interested in can be denoted as
\begin{equation} \label{OpstiUkupniPathIntegral}
Z_{G+M} = \cN \int \cD g \int \cD \varphi \, e^{iS_{\rm tot}[g,\varphi]}\,,
\end{equation}
where $S_{\rm tot}[g,\varphi]$ is the sum of the Einstein-Hilbert action and the action for the scalar field in curved spacetime,
$$
S_{\rm tot}[g,\varphi] = -\frac{1}{16\pi l_p^2} \int_{\cM_4} \rmd^4x \sqrt{-g}\, R(g) + \frac{1}{2} \int_{\cM_4} \rmd^4x \sqrt{-g} \left[ g^{\mu\nu} (\del_{\mu}\varphi)(\del_{\nu}\varphi) + m^2\varphi^2 + U(\varphi) \right]\,,
$$
where $U(\varphi)$ is a self-interaction potential of the scalar field. The corresponding lattice version of this action is given as
\begin{equation} \label{UkupnoRQGdejstvo}
\begin{array}{lcl}
S_{\rm tot}[l,\varphi] & = & \ds -\frac{1}{8\pi l_p^2} \sum_{\Delta\in T(\cM_4)} A_{\Delta}(l) \delta_{\Delta}(l) + \\
 & & \ds + \frac{1}{2} \sum_{\sigma\in T(\cM_4)} {}^{(4)}V_{\sigma}(l) g_{(\sigma)}^{\mu\nu}(l) \del\varphi_{\mu} \del\varphi_{\nu} 
 + \frac{1}{2} \sum_{v\in T(\cM_4)} {}^{(4)}V^*_v(l) \left[ m^2 \varphi_v^2 + U(\varphi_v) \right]\,. \\
\end{array}
\end{equation}
Here, a value of the scalar field $\varphi_v \in \mathbb R$ is assigned to each vertex $v \in T(\mathcal{M}_4)$. Given any $4$-simplex $\sigma\in T(\cM_4)$, one can label its five vertices as $0,1,2,3,4$, and then define a skew-coordinate system taking the vertex $4$ as the origin and edges $4-0$, $4-1$, $4-2$, $4-3$, respectively as coordinate lines for coordinates $x^{\mu}$, $\mu\in\{ 0,1,2,3 \}$. In these coordinates, the derivative $\del_{\mu}\varphi$ is replaced by the finite difference between the values of the field at the vertex $v=\mu$ and at the coordinate origin of the $4$-simplex $\sigma$ (divided by the distance between them),
$$
\del\varphi_{\mu} \equiv \frac{\varphi_{\mu} - \varphi_4}{l_{\mu 4}}\,.
$$
In addition, the metric tensor between vertices $\mu$ and $\nu$ is given in terms of edge lengths as
$$
g^{(\sigma)}_{\mu\nu}(l) \equiv \frac{l^2_{\mu 4} + l^2_{\nu 4} - l^2_{\mu\nu} }{2 l_{\mu 4} l_{\nu 4}}\,,
$$
while $g_{(\sigma)}^{\mu\nu}(l)$ is its inverse matrix. Finally, ${}^{(4)}V^*_v(l)$ is the $4$-volume of the $4$-cell surrounding the vertex $v$ in the Poincar\'e dual lattice of the triangulation $T(\cM_4)$.

After we have defined the classical action on $T(\cM_4)$, we finally proceed to define the path integral (\ref{OpstiUkupniPathIntegral}) as the state sum:
\begin{equation} \label{RQGukupnaStateSumaDef}
Z_{G+M} = \cN \int \prod_{\epsilon\in T(\cM_4)} \rmd l_{\epsilon} \, \mu(l) \int \prod_{v\in T(\cM_4)} \rmd\varphi_v \, e^{iS_{\rm tot}[l,\varphi]}\,.
\end{equation}
Here, the domain of integration for the scalar field is the Cartesian product $\realni^V$, where $V$ is the total number of vertices in the triangulation.

The state sum (\ref{RQGukupnaStateSumaDef}) defines one concrete QG model, called the Regge quantum gravity model. While it goes without saying that this is just a toy model, it is nevertheless a realistic one, since it is finite and has a correct semiclassical continuum limit (see \cite{Mikovic} for proofs). Therefore it can be used to study various aspects of quantum gravity, including the entanglement between gravity and matter fields, as we discuss next.

\subsection{\label{SubSecCalculationOfTrRhoMsq}Calculation of the trace formula}

Having formulated the Regge quantum gravity model and having the state sum (\ref{RQGukupnaStateSumaDef}) in hand, we can proceed to study the entanglement between gravity and matter, in particular by evaluating the expression for the trace of $\hat{\rho}^2_M$ given by equation (\ref{TrRhoSqPrekoZ}). In order to evaluate it, we first need to formulate the Hartle-Hawking state (\ref{DefinicijaHartleHawkingStanja}) in the framework of Regge quantum gravity model, then work out the matrix elements of the reduced density matrix (\ref{DefinicijaZ}), and finally plug them into (\ref{TrRhoSqPrekoZ}) to obtain a number. If this number is different from $1$, we can conclude that the Hartle-Hawking state features entanglement between the gravitational and matter fields.

We begin by formulating the Hartle-Hawking state (\ref{DefinicijaHartleHawkingStanja}). Consider a $4$-manifold $\cM_4$ with a nontrivial boundary $\Sigma_3=\del\cM_4$, such that the triangulation $T(\cM_4)$ induces a triangulation $T(\Sigma_3)$ on the boundary. In this sense we can distinguish the vertices, edges, areas, and tetrahedra which belong to the boundary triangulation $T(\Sigma_3)$ (from now on shortly called ``boundary'', and denoted as $\del T$), from the vertices, edges, areas, tetrahedra and $4$-simplices belonging to $T(\cM_4)$ but not to $T(\Sigma_3)$ (from now on shortly called ``bulk'', and denoted as $T$). Since the Regge quantum gravity model encodes gravitational degrees of freedom as lengths of the edges, and matter degrees of freedom as real numbers attached to vertices, we can easily split them into boundary variables $l_{\epsilon},\varphi_v$ and bulk variables $L_{\epsilon},\Phi_v$, where we maintain our previous convention to denote the bulk variables with capital letters and boundary variables with lowercase letters.

Given the bulk and the boundary, we use the formulation of the Regge quantum gravity state sum (\ref{RQGukupnaStateSumaDef}) to write down the Hartle-Hawking wavefunction as
\begin{equation} \label{RQGdefHHstanja}
\Psi_{\rm HH} [l,\varphi] = \cN \int \prod_{\epsilon\in T} \rmd L_{\epsilon} \, \mu(l,L) \int \prod_{v\in T} \rmd\Phi_v \, e^{iS_{\rm tot}[l,\varphi,L,\Phi]}\,.
\end{equation}
Next we want to construct the matrix elements of the reduced density matrix (\ref{DefinicijaZ}). To this end, we need two copies of the Hartle-Hawking state: one with matter fields $\varphi_v$ on the boundary $\del T$ of the bulk $T$, and the other with matter fields $\varphi'_v$ on the boundary $\del T$ of the bulk $\bar{T}$ defined as the mirror-reflection of $T$ with respect to the boundary $\del T$. This mirror-reflection gives rise to an additional overall minus sign in the action (\ref{UkupnoRQGdejstvo}) which is then cancelled by the complex conjugation of the imaginary unit in the exponent of the second Hartle-Hawking wavefunction in (\ref{DefinicijaZ}). Integrating over the boundary edge lengths, we end up with:
\begin{equation} \label{RQGdefinicijaZ}
Z[\varphi,\varphi'] = |\cN|^2 \int \prod_{\epsilon\in T\cup\bar{T}\cup \del T} \rmd L_{\epsilon} \, \mu(L) \int \prod_{v\in T\cup \bar{T}} \rmd\Phi_v \, e^{iS_{\rm tot}[\varphi,\varphi',L,\Phi]}\,.
\end{equation}
Note that all edge lengths are being integrated over in the ``total'' triangulation $T\cup \bar{T}\cup \del T$ (and we have thus denoted them all with a capital letter $L$ for simplicity). In contrast, the scalar field is being integrated only over the two bulks $T\cup \bar{T}$, while the boundary scalar field values $\varphi,\varphi'$ remain fixed on two identical copies of the boundary $\del T$. Also, note that
$$
S_{\rm tot}[\varphi,\varphi',L,\Phi] \equiv S_{\rm tot}[\varphi,L,\Phi] \Big|_{T\cup\del T} + S_{\rm tot}[\varphi',L,\Phi] \Big|_{\bar{T}\cup \del T}\,,
$$
where the boundary edge lengths $l$ have been relabelled as $L$ and reabsorbed into the set of bulk edge lengths.

The next step one should perform is to take the trace of (\ref{RQGdefinicijaZ}) and equate it to $1$ as in (\ref{NormalizacijaZ}), in order to make sure that the Hartle-Hawking wavefunction (\ref{RQGdefHHstanja}) is properly normalised. This leads to the equation
$$
|\cN|^2 \int \prod_{\epsilon\in T\cup\bar{T}\cup \del T} \rmd L_{\epsilon} \, \mu(L) \int \prod_{v\in T\cup \bar{T}\cup \del T} \rmd\Phi_v \, e^{iS_{\rm tot}[L,\Phi]} = 1\,,
$$
which determines the normalisation constant $\cN$ up to an overall phase factor. Note that the boundary scalar fields $\varphi$ have been integrated over and consequently reabsorbed into the bulk variables $\Phi$, similarly to edge lengths $L$. Both the integration over $L$ and the integration over $\Phi$ is now being performed over the ``total'' triangulation $T\cup\bar{T}\cup\del T$ which has no boundary.

As the final step of the construction of the trace formula (\ref{TrRhoSqPrekoZ}), we substitute (\ref{RQGdefinicijaZ}) and $\cN$ into it, to obtain:
\begin{equation} \label{RQGkonacnaTraceFormula}
\Tr_M \hat{\rho}^2_M = \frac{ \ds \int \prod_{v\in\del T} \rmd\varphi_v \int \prod_{v\in\del T} \rmd\varphi'_v \left| \int \prod_{\epsilon\in T\cup\bar{T}\cup \del T} \rmd L_{\epsilon} \, \mu(L) \int \prod_{v\in T\cup \bar{T}} \rmd\Phi_v \, e^{iS_{\rm tot}[\varphi,\varphi',L,\Phi]} \right|^2}{\ds\left( \int \prod_{\epsilon\in T\cup\bar{T}\cup \del T} \rmd L_{\epsilon} \, \mu(L) \int \prod_{v\in T\cup \bar{T}\cup \del T} \rmd\Phi_v \, e^{iS_{\rm tot}[L,\Phi]} \right)^2}\,.
\end{equation}
This is the final expression we set out to derive. It represents a concrete realisation of the trace formula (\ref{TrRhoSqPrekoZ}), it is completely well defined, and can in principle be evaluated. In practice, though, for a generic choice of the triangulation, this expression is very hard to evaluate even numerically. Therefore, in what follows we shall enforce some very hard approximations in order to make it more manageable for study. Nevertheless, by looking at the structure of the numerator and the denominator, one can already see that the two expressions can be equal to each other only in some very special cases, if at all. However, the dependence of the action $S_{\rm tot}$ on the boundary and bulk variables is such that one cannot rely on any special mathematical properties of the action which could help make the final result be $1$, for a generic choice of the spacetime triangulation. In this sense, we can conjecture already at this level that in generic cases we have
$$
\Tr_M \hat{\rho}^2_M < 1\,,
$$
as we wanted to demonstrate.

But in order to give a more convincing argument, let us study a special case and try to evaluate this trace to the very end. The simplest possible example of a triangulation $T$ is a single $4$-simplex. Labelling its vertices as $1,2,3,4,5$, we can depict it with a following diagram:
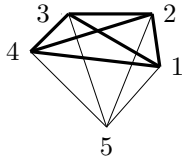
\begin{center}
\begin{tikzpicture}
\draw[very thick] (0,2) -- (1.6,2.5) ;
\draw[very thick] (0,2) -- (1.7,1.8) ;
\draw[very thick] (0,2) -- (0.5,2.5) ;
\draw[very thick] (1.7,1.8) -- (0.5,2.5) ;
\draw[very thick] (1.7,1.8) -- (1.6,2.5) ;
\draw[very thick] (0.5,2.5) -- (1.6,2.5) ;
\draw[very thin] (1,1) -- (0,2) ;
\draw[very thin] (1,1) -- (1.6,2.5) ;
\draw[very thin] (1,1) -- (0.5,2.5) ;
\draw[very thin] (1,1) -- (1.7,1.8) ;
\filldraw[black] (1.7,1.8) circle (0pt) node[anchor=west] {\footnotesize 1};
\filldraw[black] (1.6,2.5) circle (0pt) node[anchor=west] {\footnotesize 2};
\filldraw[black] (0.4,2.5) circle (0pt) node[anchor=east] {\footnotesize 3};
\filldraw[black] (0,2) circle (0pt) node[anchor=east] {\footnotesize 4};
\filldraw[black] (1,1) circle (0pt) node[anchor=north] {\footnotesize 5};
\end{tikzpicture}
\end{center}
The $4$-simplex has five boundary tetrahedra, namely
$$
\tau_{1234}\,,\qquad
\tau_{1235}\,,\qquad
\tau_{1245}\,,\qquad
\tau_{1345}\,,\qquad
\tau_{2345}\,.
$$
The first tetrahedron, $\tau_{1234}$, is depicted with thick edges, and we will choose it to be the boundary $\del T$. Since we do not want the four remaining tetrahedra to belong to the boundary, we will glue them onto each other in pairs, as
$$
\tau_{1235} \equiv \tau_{1245}\,,\qquad \tau_{1345} \equiv \tau_{2345}\,.
$$
This means that every point in $\tau_{1235}$ is identified with the corresponding point in $\tau_{1245}$, and similarly with the other pair of tetrahedra. In this way we obtain a manifold with a nontrivial topology, but described with only five vertices and one boundary tetrahedron. In order for this gluing to be consistent, the gravitational and matter degrees of freedom living on $T\cup\del T$ must satisfy the following constraints:
\begin{equation} \label{VezeZaPrviSimpleks}
l_{14} = l_{23} = l_{24} = l_{13}\,, \qquad L_{25} = L_{15}\,, \qquad L_{45} = L_{35}\,, \qquad \varphi_2 = \varphi_1\,, \qquad \varphi_4 = \varphi_3\,.
\end{equation}
This leaves us with the following independent degrees of freedom living on the $4$-simplex:
$$
l_{12}\,,\qquad
l_{13}\,,\qquad
L_{15}\,,\qquad
l_{34}\,,\qquad
L_{35}\,,\qquad
\varphi_1\,,\qquad
\varphi_3\,,\qquad
\Phi_5\,,
$$
where we have denoted the bulk degrees of freedom with capital letters and boundary degrees of freedom with lowercase letters. The $4$-simplex diagram above is the graphical representation of the Hartle-Hawking wavefunction $\Psi_{\rm HH} [l,\varphi]$ (\ref{RQGdefHHstanja}).

Next we construct $\bar{T}$. Since the boundary tetrahedron $\del T$ defines a single $3$-dimensional hypersurface, there is precisely one axis in $4$-dimensional space which is orthogonal to $\del T$. Performing the reflection of $T$ with respect to $\del T$ is therefore identical to reversing the orientation of this orthogonal axis. In this way we construct another $4$-simplex, with vertices labeled $1,2,3,4,6$ and depicted as
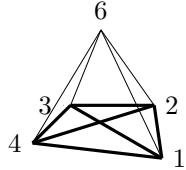
\begin{center}
\begin{tikzpicture}
\draw[very thick] (0,2) -- (1.6,2.5) ;
\draw[very thick] (0,2) -- (1.7,1.8) ;
\draw[very thick] (0,2) -- (0.5,2.5) ;
\draw[very thick] (1.7,1.8) -- (0.5,2.5) ;
\draw[very thick] (1.7,1.8) -- (1.6,2.5) ;
\draw[very thick] (0.5,2.5) -- (1.6,2.5) ;
\draw[very thin] (0.9,3.5) -- (0,2) ;
\draw[very thin] (0.9,3.5) -- (1.6,2.5) ;
\draw[very thin] (0.9,3.5) -- (0.5,2.5) ;
\draw[very thin] (0.9,3.5) -- (1.7,1.8) ;
\filldraw[black] (1.7,1.8) circle (0pt) node[anchor=west] {\footnotesize 1};
\filldraw[black] (1.6,2.5) circle (0pt) node[anchor=west] {\footnotesize 2};
\filldraw[black] (0.4,2.5) circle (0pt) node[anchor=east] {\footnotesize 3};
\filldraw[black] (0,2) circle (0pt) node[anchor=east] {\footnotesize 4};
\filldraw[black] (0.9,3.5) circle (0pt) node[anchor=south] {\footnotesize 6};
\end{tikzpicture}
\end{center}
One can see that the main difference between the $4$-simplex $\sigma_{12346}$ and the previously constructed $4$-simplex $\sigma_{12345}$ is that the vertex $6$ is on the ``opposite side'' of the tetrahedron $\tau_{1234}$ as compared to the vertex $5$ of $\sigma_{12345}$.

Like we did for $\sigma_{12345}$, we again want to glue the boundary tetrahedra pairwise, so that only the tetrahedron $\tau_{1234}$ remains as the boundary $\del \bar{T}$. The pairwise gluing of tetrahedra
$$
\tau_{1236} \equiv \tau_{1246}\,,\qquad \tau_{1346} \equiv \tau_{2346}\,
$$
gives rise to the constraints
$$
l_{14} = l_{23} = l_{24} = l_{13}\,, \qquad L_{26} = L_{16}\,, \qquad L_{46} = L_{36}\,, \qquad \varphi'_2 = \varphi'_1\,, \qquad \varphi'_4 = \varphi'_3\,,
$$
where only the constraints containing the vertex $6$ are additional to (\ref{VezeZaPrviSimpleks}). This leaves us with the following independent degrees of freedom living on $\sigma_{12346}$:
$$
l_{12}\,,\qquad
l_{13}\,,\qquad
L_{16}\,,\qquad
l_{34}\,,\qquad
L_{36}\,,\qquad
\varphi'_1\,,\qquad
\varphi'_3\,,\qquad
\Phi_6\,.
$$
As noted in the general discussion leading to equation (\ref{RQGkonacnaTraceFormula}), the matter degrees of freedom on the boundary of $T$ are different than the corresponding degrees of freedom living on the boundary of $\bar{T}$, despite the fact that the boundary is identical, $\del \bar{T} \equiv \del T$. To that end, we have added a prime to $\varphi$ in the above equations. Like for the $4$-simplex $\sigma_{12345}$, the diagram of the $4$-simplex $\sigma_{12346}$ above is the graphical representation of the (complex-conjugate) Hartle-Hawking wavefunction $\Psi^*_{\rm HH} [l,\varphi']$.

At this point we are ready to glue $T$ and $\bar{T}$ along the common boundary $\del T$, to obtain the manifold $T\cup \bar{T} \cup \del T$ which has no boundary. It is depicted on the diagram below.
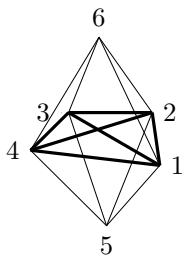
\begin{center}
\begin{tikzpicture}
\draw[very thick] (0,2) -- (1.6,2.5) ;
\draw[very thick] (0,2) -- (1.7,1.8) ;
\draw[very thick] (0,2) -- (0.5,2.5) ;
\draw[very thick] (1.7,1.8) -- (0.5,2.5) ;
\draw[very thick] (1.7,1.8) -- (1.6,2.5) ;
\draw[very thick] (0.5,2.5) -- (1.6,2.5) ;
\draw[very thin] (1,1) -- (0,2) ;
\draw[very thin] (1,1) -- (1.6,2.5) ;
\draw[very thin] (1,1) -- (0.5,2.5) ;
\draw[very thin] (1,1) -- (1.7,1.8) ;
\draw[very thin] (0.9,3.5) -- (0,2) ;
\draw[very thin] (0.9,3.5) -- (1.6,2.5) ;
\draw[very thin] (0.9,3.5) -- (0.5,2.5) ;
\draw[very thin] (0.9,3.5) -- (1.7,1.8) ;
\filldraw[black] (1.7,1.8) circle (0pt) node[anchor=west] {\footnotesize 1};
\filldraw[black] (1.6,2.5) circle (0pt) node[anchor=west] {\footnotesize 2};
\filldraw[black] (0.4,2.5) circle (0pt) node[anchor=east] {\footnotesize 3};
\filldraw[black] (0,2) circle (0pt) node[anchor=east] {\footnotesize 4};
\filldraw[black] (1,1) circle (0pt) node[anchor=north] {\footnotesize 5};
\filldraw[black] (0.9,3.5) circle (0pt) node[anchor=south] {\footnotesize 6};
\end{tikzpicture}
\end{center}
It consists of two $4$-simplices $\sigma_{12345}$ and $\sigma_{12346}$ constructed above and glued along the common tetrahedron $\tau_{1234}$. The full set of independent gravitational degrees of freedom is
$$
l_{12}\,,\qquad
l_{13}\,,\qquad
l_{34}\,,\qquad
L_{15}\,,\qquad
L_{16}\,,\qquad
L_{35}\,,\qquad
L_{36}\,,
$$
while the independent matter degrees of freedom are
$$
\varphi_1\,,\qquad
\varphi_3\,,\qquad
\varphi'_1\,,\qquad
\varphi'_3\,,\qquad
\Phi_5\,,\qquad
\Phi_6\,.
$$
This diagram is the graphical representation for the matrix element $Z[\varphi,\varphi']$ of the reduced density matrix $\hat{\rho}_M$ (see equations (\ref{RQGdefinicijaZ}) and (\ref{DefinicijaZ})).

Applying the general trace formula (\ref{RQGkonacnaTraceFormula}) to our case then gives
\begin{equation} \label{RQGkonacnaTraceFormulaPrimer}
\Tr_M \hat{\rho}^2_M = \frac{ \ds \int \rmd\varphi_1 \rmd\varphi_3 \rmd\varphi'_1 \rmd\varphi'_3 \left| \int \rmd^7L \, \mu(L) \int \rmd\Phi_5 \rmd\Phi_6 \, e^{iS_{\rm tot}[\varphi,\varphi',L,\Phi]} \right|^2}{\ds\left( \int \rmd^7L \, \mu(L) \int \rmd^4\Phi \, e^{iS_{\rm tot}[L,\Phi]} \right)^2}\,,
\end{equation}
where
$$
\rmd^7L \equiv \rmd l_{12} \rmd l_{13} \rmd l_{34} \rmd L_{15} \rmd L_{16} \rmd L_{35} \rmd L_{36}\,,
$$
and
$$
\rmd^4\Phi \equiv \rmd\varphi_1 \rmd\varphi_3 \rmd\Phi_5 \rmd\Phi_6\,.
$$
Note that the action in the denominator is evaluated using $\varphi'_1=\varphi_1$ and $\varphi'_3=\varphi_3$, as explained in the general discussion above. In order to make the equation (\ref{RQGkonacnaTraceFormulaPrimer}) fully explicit, we need to choose the values of the free parameters in the classical action (\ref{UkupnoRQGdejstvo}) and the measure (\ref{IzabranaRQGmera}). The parameters of the action are the Planck length $l_p$, the mass $m$ of the scalar field, and the self-interaction potential $U(\varphi)$. For the purpose of this example, the simplest possible choice is the free massless scalar field, so that we have
$$
l_p = 10^{-35}\,{\rm m}\,, \qquad m=0, \qquad U(\varphi) = 0\,.
$$
Second, the measure (\ref{IzabranaRQGmera}) contains a single free parameter $L_{\mu}$. This parameter can be connected to the value of the effective cosmological constant $\Lambda$, via the relation
$$
\Lambda = \frac{l_p^2}{2L_{\mu}^4}\,,
$$
see \cite{Mikovic,MikovicVojinovicCCletter,MikovicVojinovicCCveliki} for details. Taking the observed value $\Lambda = 10^{-52} \,{\rm m}^{-2}$ (also often quoted as a dimensionless product $\Lambda l_p^2 = 10^{-122}$), we obtain
$$
L_{\mu} = 10^{-5}\,{\rm m}\,.
$$
Using these numeric values of the parameters, the right-hand side of (\ref{RQGkonacnaTraceFormulaPrimer}) is fully specified, and can be evaluated using a computer. However, in order to render the calculation more manageable, for the purpose of this paper we instead choose to evaluate (\ref{RQGkonacnaTraceFormulaPrimer}) with $L_{\mu} = 10^{-33}\,{\rm m}$, which corresponds to a larger cosmological constant, $\Lambda l_p^2 = 10^{-8}$, to speed up the convergence of the Monte-Carlo integration method. The result is strictly less than one,
$$
\Tr_M \hat{\rho}^2_M = 0.977 \pm 0.002\,,
$$
as we had set out to demonstrate. Note that, although close to one, the above result is: (i) strictly smaller than one (within the computational error); (ii) obtained within extremely simplified toy model whose system consists of only two $4$-simplices of spacetime. Thus, our result can serve as a proof of principle that gravity-matter entanglement is always present. The total amount of such entanglement in realistic models, as well as its spatial distribution, remains to be further explored. Namely, note that even though the approximation of product gravity-matter states has been up to now successfully applied, the {\em overall} entanglement between the two systems, considered within complex realistic situations/models, does not at all have to be small, nor its effects negligible. Indeed, the standard entanglement that is considered to cause the decoherence of matter by the environment and the quantum-to-classical transition has profoundly striking effects, despite the fact of being difficult to characterise, evaluate and manipulate.

\section{\label{SecConclusions}Conclusions}


\subsection{Summary of the results}

We analyse the quantum gravity coupled to the most common matter fields (namely, scalar, spinor and vector fields), and show that the gravity and matter are generically entangled, as a consequence of the nonseparability of the scalar constraint $\cC$, and in some cases the $3$-diffeo constraint $\cC^M_i$. Thus, simple separable gravity-matter product states are excluded from the physical Hilbert space, unless the constraint equations feature some deep unknown property which allows for the invariance of a whole class of product states. We demonstrate this in two different ways: (i) within the general abstract nonperturbative canonical formalism, by directly analysing the mathematical structure of the constraints, and (ii) within the path integral formalism, by directly checking for entanglement of the Hartle-Hawking state in the Regge model of quantum gravity.


\subsection{Discussion of the results}

This {\em gauge-protected} decoherence due to the entanglement (in contrast to the standard {\em ``for all practical purposes''} dynamical one) offers a possibly deeper fundamental explanation of the long-standing problem of the quantum-to-classical transition: the matter does not {\em decohere}, it is by default {\em decohered}. 

Any potential entanglement, either dynamical or gauge-protected one, depends on the details of the coupling between matter and gravity. For the purpose of this paper, the coupling is prescribed by the strong equivalence principle, which states that the equations of motion for all matter fields must locally be identical to the equations of motion for those fields in flat spacetime. This is implemented by choosing the action for matter fields with minimal coupling prescription, and employed in both the canonical and the path integral frameworks. We should stress that the validity of the strong equivalence principle is a sufficient, but potentially not a necessary assumption for our main result. Namely, it is plausible that nonminimal coupling choices, involving explicit spacetime curvature terms in the matter Lagrangian, could also lead to the conclusion that entanglement between gravity and matter is unavoidable. However, it is also possible that one could come up with some particular complicated choice of nonminimal coupling which does admit some nonentagled states. In order to avoid complicating the analysis with such cases, given that nonminimal coupling between gravity and matter has absolutely no experimental evidence in its favor so far, we have chosen to assume the validity of the strong equivalence principle throughout the paper.

In standard QM entanglement is a generic consequence of the interaction. Nevertheless, there exist alternative mechanisms for creating it, such as the indistinguishability of identical particles, leading to effective ``exchange interactions''. This new gauge-protected gravity-matter entanglement can thus introduce additional ``effective interaction'', which can possibly result in corrections to Einstein's weak equivalence principle (see for example \cite{PipaPaunkovicVojinovic}).

It is interesting to note that a possible peculiar impact of the quantised gravity to the whole decoherence programme was already inferred in Zurek's seminal paper~\cite{zur:81}, where on page 1520 the author writes: (the assumption of pairwise interactions) ``is customary and clear, even though it may prevent one from even an approximate treatment of the gravitational interaction beyond its Newtonian pairwise form.'' Our result confirms Zurek's disclaimer -- gravity (environment $\mathcal{E}$) is generically entangled with the {\em whole} matter (both the system $\mathcal{S}$ and the apparatus $\mathcal{A}$), that way allowing for non-trivial tripartite system-apparatus-environment {\em effective} interaction of the form $\mathcal{H}_{\mathcal{S}\mathcal{A}\mathcal{E}}$, explicitly excluded in~\cite{zur:81}. In other words, the environment (spacetime) interaction with the matter could potentially disturb the system-apparatus correlations, thus violating the stability criterion of a faithful measurement (see~\cite{sch:05}, p.~1271). 

As a consequence of generic gravity-matter entanglement, the effective interaction between gravity and matter forbids the existence of a single background spacetime. Thus, when concerning quantum effects of gravity, one cannot talk of ``matter in a point of space'', confirming the conjecture that spacetime is an ``emergent phenomenon''. In contrast to this, Penrose argues that spacetime, seen as a (four-dimensional) differentiable manifold, does not support superpositions of massive bodies and the corresponding (relative) states of gravity (i.e., the gravity-matter entanglement), leading to the objective collapse onto the product states of matter and (classical) spacetime~\cite{pen:96}. Our result can therefore be treated as a possible criterion for a plausible candidate theory of quantum gravity. 

Finally, not allowing product states between the matter and gravity is in tune with the relational approach to physics~\cite{rov:04}, in particular to quantum gravity (note that the original name for the many-world interpretation of QM was the {\em ``Relative State'' Formulation of Quantum Mechanics}~\cite{eve:57}). See also \cite{DeWittbook} for an interesting treatment of relative state and decoherence approaches.


\subsection{Relation to common quantum gravity research programs}

In order to discuss our results in the context of various quantum gravity research programs, note that the gauge-protected entanglement between gravity and matter should exist in any model of quantum gravity with matter which respects local Poincar\'e symmetry. In this sense, various approaches to quantum gravity can be classified into four distinct categories.
\begin{itemize}
\item[(i)] The first category represents models which explicitly respect (or at least aim to respect) local Poincar\'e symmetry. These include nonperturbative string theory/M-theory \cite{Duff,Lambert,BeckerBeckerSchwarz}, loop quantum gravity \cite{rov:04,rov:vid:14}, Wheeler-DeWitt quantization \cite{DeWitt,ADM}, and similar approaches.
\item[(ii)] The second category represents models in which local Poincar\'e symmetry is explicitly broken. These include perturbative quantum gravity \cite{FeynmanLecturesOnGravity}, petrurbative string theory \cite{BeckerBeckerSchwarz}, causal dynamical triangulations approach \cite{Loll}, doubly-special relativity models \cite{DSRreview}, Ho\v rava-Lifshitz gravity \cite{Horava}, various nonrelativistic quantization proposals, and so on.
\item[(iii)] The third category represents models in which it is not clear whether local Poincar\'e symmetry is broken or not. For example, in the asymptotic safety approach \cite{AsympSafetyReview} this may depend on the properties of the fixed point. In noncommutative geometry \cite{Connes,Sakellariadou} it depends on the particular choice of the algebra. In higher-derivative theories and theories with propagating torsion \cite{Hehl} it may depend on various details of the model, etc.
\item[(iv)] Finally, the fourth category represents models which have not been developed enough to allow for coupling of matter fields. In models like entropic gravity \cite{Verlinde,Jacobson} and causal set theory \cite{Henson,Sorkin}, it is not obvious how to couple matter fields to gravity, and whether this coupling would violate local Poincar\'e invariance or not.
\end{itemize}

It should be clear that our results apply to the first category of quantum gravity models, while for other three categories it either does not apply, or it is an open question. We should also state that the validity of local Poincar\'e symmetry is ultimately an experimental question, one over which various quantum gravity proposals may disagree.

In relation to the previous comment, it is worthwhile to also discuss the impact of possible anomalies to the gauge protected entanglement. As we have discussed in the final paragraph of section \ref{SecGravConstraint}, the entanglement is a consequence of the scalar constraint $\hat{\cC}$, see (\ref{QuantumScalarConstraintEquation}), and for certain types of matter fields also of the $3$-diffeo constraint $\hat{\cC}_i$ in (\ref{QuantumKinematicConstraintEquations}), while the local Lorentz constraint $\hat{\cC}_{ab}$ in (\ref{QuantumKinematicConstraintEquations}) does not require entanglement. From this one can see that if the theory features anomalies due to the breaking of the $4D$ diffeomorphism symmetry, one cannot impose $\hat{\cC}$ and $\hat{\cC}_i$ as the Gupta-Bleuler-like conditions on the Hilbert space of the theory, and thus all subsequent results regarding the entanglement are void. In short, there cannot be any gauge protected entanglement if there is no relevant gauge symmetry to begin with. Nevertheless, if the theory features anomalies due to the breaking of the local Lorentz or any internal symmetries, while maintaining diffeomoprhism symmetry at the quantum level, the gauge protected entanglement will not be influenced by the anomaly.


\subsection{Future lines of research}

One of the main lines of future work would be to perform a detailed numerical analysis of $\Tr \hat\rho^2_M$ and the von Neumann entropy $S(\hat\rho_M)$ for the Hartle-Hawking state (either within the Regge, or some other QG model). The latter quantity, called the {\em entropy of entanglement}, represents the measure of the entanglement in {\em pure} and {\em bipartite} states~\cite{ben:ber:pop:sch:96}, in our case between gravity and matter in the Hartle-Hawking state. The precise numerical deviation of the $\Tr \hat\rho^2_M$ from its maximal value 1 could indicate in which cases this new entanglement has relevant physical consequences. This way, it would be possible to determine the boundaries of validity of the assumption of the product gravity-matter states of the form $\ket G \ket M$, which has been up to now used in numerous studies (analogously to the case of determining the regimes in which two coherent states become effectively orthogonal). In connection to this, one could analyse in more detail quantitatively to what extent the gauge-protected gravity-matter entanglement constrains the existence of macroscopic superpositions, and its effect to the quantum-to-classical transition (see the related work~\cite{BruknerNatureCom,BruknerNaturePhys,per:ros:64,kay:98,bay:oza:09,bru:15}).

Further, studying the structure of the gauge-imposed entanglement for a tripartite system of gravity-matter-EM fields might bring qualitatively new effects. Unlike the case of pure bipartite states, where any two entangled states could be obtained from each other by Local Operations and Classical Communication (LOCC), thus forming a single class of entangled states and providing a unique measure of entanglement, the multipartite entanglement has a more complex structure. Indeed, in the tripartite case, in addition to the trivial classes of purely bipartite entanglement, say, $\ket a (\ket{b_1 c_1} + \ket{b_2 c_2})$, genuine tripartite entanglement consists of a number of inequivalent classes of entangled states: in the simplest case of three qubits we have two classes of tripartite entanglement, represented by the states $\ket{GHZ} = (\ket{000} + \ket{111})/\sqrt 2$ and $\ket W = (\ket{001} + \ket{010} + \ket{100})/\sqrt 3$, which cannot be obtained from each other by the means of LOCC, but as soon as neither of the subsystems is a qubit, there exist infinitely many inequivalent classes~\cite{dur:vid:cir:00}. 

It would also be interesting to see how other QG candidates incorporate the general gravity constraints regarding the entanglement with matter, in particular the string theory. Namely, perturbative string theory is formulated by manifestly breaking the gauge symmetry (a consequence of perturbative expansion of the gravitational field). The existence of the gravity-matter entanglement in, say Hartle-Hawking state, would then present a strong argument that the gauge symmetry can be restored in a tentative nonperturbative formulation of string theory. In connection to this, one could analyse the entanglement between different spacetime regions induced by the gauge-protected gravity-matter entanglement, and compare it to that present in theories based on the AdS/CFT correspondence and the holographic principle~\cite{don:16,raa:16}. Namely, entanglement is a property of a quantum state {\em with respect} to a particular factorisation of a composite system into its factor sub-systems. To illustrate this, consider a particle in a two-dimensional plane. Given orthogonal axes $x$ and $y$ of a 2D plane, the Hilbert space of the system is given by $\cH = \cH_x \otimes \cH_y$, and the equal spatial superposition (for simplicity, we omit the overall normalisation constant) $\ket\varphi \sim (\ket{a}_x + \ket{b}_x)\ket{0}_y$, with $a,b\in\mathbb{R}$, is clearly separable, with respect to the given factorisation of $\cH$. Nevertheless, with respect to {\em any other} factorisation of $\cH$, defined by any other axes $X$ and $Y$ inducing the Hilbert-space factorisation $\cH = \cH_X \otimes \cH_Y$, the system is entangled. As an example, for axes $X$ and $Y$ obtained by rotating $x$ and $y$ by $-\pi/4$, the same state of the system is maximally entangled, $\ket\varphi \sim (\ket{a/\sqrt{2}}_X\ket{a/\sqrt{2}}_Y + \ket{b/\sqrt{2}}_X\ket{b/\sqrt{2}}_Y)$ (for the entanglement in the second quantisation formalism, and its dependence on the choice of fundamental modes, see for example~\cite{ved:03}). Following the above example, one might expect that the existence of the entanglement between gravity and matter would induce the entanglement between two generic spacetime regions (each containing a portion of both gravitational and matter degrees of freedom). Possible relationship between this, gauge-protected entanglement, and that present as a consequence of assumptions that do not explicitly rely on the existence of local Poincar\'e symmetry (holography and the AdS/CFT correspondence) would indicate interesting fundamental connections that could help breaching the long-standing gap between quantum mechanics and general relativity.

Finally, detecting gravity-matter entanglement in the experiment might not be that far from the reach of the current or the near-future technology, see~\cite{pfi:etal:16} for a recent proposal of testing gravitational decoherence. Proposing, and possibly performing, experiments to distinguish different contributions of the gravitational interaction to the decoherence of matter, in particular the generic one based on the gauge symmetry constraints, presents a relevant direction of further research.

\bigskip

\centerline{\bf Acknowledgments}

\bigskip

We would like to thank Rafael Sorkin for helpful suggestions and comments.

MV was supported by the project ON171031 of the Ministry of Education, Science and Technological Development of the Republic of Serbia, and the bilateral scientific cooperation between Portugal and Serbia through the project ``Quantum Gravity and Quantum Integrable Models - 2015-2016'', no. 451-03-01765/2014-09/24 supported by the Foundation for Science and Technology (FCT), Portugal, and the Ministry of Education, Science and Technological Development of the Republic of Serbia. NP acknowledges the support of SQIG -- Security and Quantum Information Group, the IT project QbigD funded by FCT PEst-OE/EEI/LA0008/2013 and UID/EEA/50008/2013 and the bilateral scientific cooperation between Portugal and Serbia through the project ``Noise and measurement errors in multi-party quantum security protocols'', no. 451-03-01765/2014-09/04 supported by the Foundation for Science and Technology (FCT), Portugal, and the Ministry of Education, Science and Technological Development of the Republic of Serbia.

The authors would also like to thank the Erwin Schr\"odinger International Institute for Mathematics and Physics (ESI Vienna), for the warm hospitality during the workshop ``Quantum Physics and Gravity'' (29. May - 30. June 2017) and partial support.


\bibliography{gment}

\end{document}